\documentclass[10pt,journal,compsoc]{IEEEtran}
\usepackage{tabu}                      
\usepackage{booktabs}                  
\usepackage{amsmath}
\usepackage{graphicx,calc}
\usepackage[hidelinks]{hyperref}
\graphicspath{{figures/}{pictures/}{images/}{./}}
\usepackage{array}
\usepackage{textcase}
\usepackage{enumitem}
\usepackage{caption}
\usepackage{subcaption}
\usepackage{titlecaps}
\usepackage{wrapfig}
\usepackage{multirow}
\usepackage{makecell}
\usepackage{xcolor}

\AtBeginDocument{
}

\newlist{compactenum}{enumerate}{4}
\setlist[compactenum,1]{nolistsep}
\setlist{noitemsep,parsep=0pt,partopsep=0pt} 
\newlength\myheight
\newlength\mydepth
\settototalheight\myheight{Xygp}
\newcommand*\inlinegraphics[1]{%
  \settototalheight\myheight{Xygp}%
  \settodepth\mydepth{Xygp}%
  \raisebox{-\mydepth}{\includegraphics[height=\myheight]{#1}}%
}
\newcommand*\inlinegraphic[1]{%
  \settototalheight\myheight{Xygp}%
  \settodepth\mydepth{Xygp}%
  \raisebox{-0.8\mydepth}{\includegraphics[height=1.2\myheight]{#1}}%
}

\newcommand{\eg}{\textit{e.g.}}
\newcommand{\ie}{\textit{i.e.}}
\newcommand{\etcc}{\textit{etc.,}}
\newcommand{\etc}{\textit{etc\rev{.}}}
\newcommand{\etal}{\textit{et al.}}

\newcommand{\demo}{\underline{\linkhl{\url{https://emordle.github.io/}}}}
\newcommand{\survey}{\underline{\linkhl{\url{https://osf.io/dxvhz/}}}}
\newcommand{\abbr}{emordle}

\newcommand{\rev}[1]{{\color{black}{#1}}} 
\newcommand{\revv}[1]{{\color{black}{#1}}} 
\newcommand{\linkhl}[1]{{\color{blue}{#1}}} 
\newcommand{\draft}[1]{{\color{black}{#1}}}  

\ifCLASSOPTIONcompsoc
  \usepackage[nocompress]{cite}
\else
  \usepackage{cite}
\fi
\ifCLASSINFOpdf
\else
\fi
\begin{document}
%

\title{Creating Emordle: Animating Word Cloud for Emotion Expression}
%
%
%
%

\author{Liwenhan Xie, Xinhuan Shu, Jeon Cheol Su, Yun Wang, Siming Chen, and Huamin Qu
\IEEEcompsocitemizethanks{
    \IEEEcompsocthanksitem  Liwenhan Xie, Xinhuan Shu, Jeon Cheol Su, and Huamin Qu are the Hong Kong University of Science and Technology. Liwenhan Xie is also a visiting student at Fudan University.\protect\\
    E-mail: \{lxieai, xinhuan.shu, csjeon, huamin\}@cse.ust.hk.\protect
\IEEEcompsocthanksitem Yun Wang is with Microsoft Research Asia.\protect
Email: wangyun@microsoft.com.

\IEEEcompsocthanksitem Siming Chen is with Fudan University and the Shanghai Key Laboratory of Data Science. Email: simingchen@fudan.edu.cn.\protect
\IEEEcompsocthanksitem Siming Chen is the corresponding author.\\

}

\thanks{© 2023 IEEE.  This is the author’s version of the article that has been accepted in IEEE Transactions on Visualization and
Computer Graphics. Personal use of this material is permitted.  Permission from IEEE must be obtained for all other uses, in any current or future media, including reprinting/republishing this material for advertising or promotional purposes, creating new collective works, for resale or redistribution to servers or lists, or reuse of any copyrighted component of this work in other works. The final version of this record will be available at: 10.1109/TVCG.2022.3209383.}
}

%
%

\markboth{TO APPEAR IN IEEE Transactions on Visualization and Computer Graphics. DOI: 10.1109/TVCG.2023.3286392}%
{Xie \MakeLowercase{\textit{et al.}}: Creating Emordle: Animating Word Cloud for Emotion Expression}
%



\IEEEtitleabstractindextext{%
\begin{abstract}
We propose \abbr{}, a conceptual design that animates wordles (compact word clouds) to deliver their emotional context to audiences. To inform the design, we first reviewed online examples of animated texts and animated wordles, and summarized strategies for injecting emotion into the animations.
We introduced a composite approach that extends an existing animation scheme for one word to multiple words in a wordle with two global factors: the randomness of text animation (entropy) and the animation speed (speed).
To create an \abbr{}, general users can choose one predefined animated scheme that matches the intended emotion class and fine-tune the emotion intensity with the two parameters. We designed proof-of-concept emordle examples for four basic emotion classes, namely happiness, sadness, anger, and fear. We conducted two controlled crowdsourcing studies to evaluate our approach. The first study \revv{confirmed} that people generally agreed on the conveyed emotions from well-crafted animations, and the second one demonstrated that our identified factors helped fine-tune the extent of the emotion delivered. We also invited general users to create their own \abbr{}s based on our proposed framework. \revv{Through this user study, we confirmed the effectiveness of the approach. We concluded with implications for future research opportunities of supporting emotion expression in visualizations.}

\end{abstract}

}

\maketitle

\IEEEdisplaynontitleabstractindextext

%
\IEEEpeerreviewmaketitle

\IEEEraisesectionheading{\section{Introduction}\label{sec:introduction}}

\IEEEPARstart{W}{ith} an appealing appearance and simple creation process, wordle has been popular in the wild for non-analytical purposes~\cite{viegas09participatory}.
Yet, it suffers from loss of context~\cite{felix17taking}.
Due to the ambiguity of language, presenting only keywords poses challenges to sense-making.
Regarding this, various techniques have been introduced to amplify wordle's semantic context~\cite{hearst2020semantic}.
However, little scholarly attention has been paid to the emotional context, which also plays an essential role in communication~\cite{lan21kineticharts}.
For instance, ``perfect'' could be genuine praise or sarcasm, dependent on the tone.
Motivated to mend the research gap, we explored encoding the emotional context into the wordle.


Among all visual encoding channels, we investigated a specific aspect: animation.
Our inspiration was drawn from animated text (\eg, syllabus, words, and sentences), which has been widely applied in advertisements, films, and lyric videos.
A classic example is the opening for Pixar movies, where Luxo \textit{Jr.}~\cite{luxos}, the lively lamp, jumps over the elastic alphabet ``I'' which brings delight to the audience while creating a relaxed atmosphere. 
Animated text can mimic body movements or natural phenomena by changing color, size, and position.
Evidence revealed that people might perceive the intended emotions of an animated text~\cite{stone2004type,malik09communicating}.
While animated data storytelling has received growing interest recently (\eg,~\cite{Shu21Gif,li2023geocamera}), most studies have limited their scope to charts that are based on structured data (\eg~\cite{lan21kineticharts,roslingifier2022shin}), where wordle animation remains an understudied direction~\cite{Jo15WordlePlus,chi15morphable}.
With limited real-life instances of animated wordle, we analyzed online instances of animated text with multiple words or emotion designs to inform our design of emotional animated wordle, namely ``\abbr{}''.

Our work also \revv{seeks} practical approaches for generating \abbr{}, in line with the common goal in wordle tools (\eg{}, \cite{Koh10ManiWordle, wordArt, Wang18EdiWordle, Kulahcioglu20affect}) to help create satisfying results for general users under a casual context.
So far, there have been two major challenges in creating \abbr{}s.
First, it remains unclear how to deliver emotion in the animation of a wordle.
Second, even for people skilled in design, off-the-shelf animation software, such as Adobe After Effects~\cite{aftereffects}, requires intensive human labor to tweak each detailed property and arrange the timeline or keyframes with substantial functionalities and property controls.
We \revv{propose} a composite approach that reuses an existing animation design for one word and propagates it to multiple words.
Thereby, an animated wordle is a composite of words with variants of the backbone animation, coordinated by external parameters.
Our work is one of the few attempts to incorporate emotional context into visualization creation.


This paper first starts by articulating the goals of an \abbr{} generation method, \ie, supporting emotion delivery, facilitating fast generation, and enabling personalized design (\autoref{sec:goal}).
Bearing these goals, we referred to existing artifacts that were most relevant to the concept of \abbr{}.
Through content analysis on a corpus with 77 animated text clips, we derived a design space for animated text and \rev{identified} three strategies for delivering emotions in animated text design (\autoref{sec:design-space}).
Informed by the findings, we proposed to extend an emotional animated text to an \abbr{}, with external parameters of ``speed'' and ``entropy'' (\autoref{sec:method}).
The original animation scheme determines the base emotion class, and the two parameters control low-level animation details and fine-tune the intensity of the displayed emotion.
Four design cases for happiness, sadness, anger, and fear were produced and evaluated in two crowdsourced studies, which demonstrated the crowd agreement on basis emotions over particular animation schemes, and validated the influence of speed and entropy on the perceived emotion intensity (\autoref{sec:eval}).
Lastly, we applied our approach to a proof-of-concept system for creating \abbr{}s on the four emotions and conducted a user study (\autoref{sec:user-study}), where we summarized the insights and implications for future tool support in-depth (\autoref{sec:discussion}). 

In general, the contributions of this paper are threefold.
\begin{itemize}
    \item \draft{A proposal of \abbr{}---a conceptual design to convey emotion through the animation of wordles. Through a qualitative study, we derived a design space of animated text and identified strategies to deliver emotion.}
    \item A composite approach to generate ``\abbr{}''. We extracted two parameters, speed and entropy, to help fine-tune the emotion intensity. We showed how to instantiate \abbr{}s on four emotions. We also implemented an \abbr{} creation tool.

    \item Two controlled studies confirming a level of crowd agreement over the \abbr{} design and a user study gaining people's feedback on the creation tool.
\end{itemize}

\section{Related Work}
\label{sec:related-work}
Our work is closely related to animated text, visualization design for delivering emotion, and wordle variants.

\subsection{Animated Text}
Animated text, also termed \textit{kinetic typography} or \textit{moving text}, is designed to express emotional content, promote engagement, and portray compelling characters through text movement and style changes~\cite{malik09communicating, Lee02KineticTypographyEngine}.
It has become commonplace in films, advertisements, lyric videos, \etc{}, due to its intuitiveness and expressiveness. Some works have also investigated its applications in captioning~\cite{Lee07EmotiveCaptioning} and instant messaging~\cite{Forlizzi03Kinedit, Wang04communicate, Lee06Kinetic}.
A case study~\cite{Suguru96Multiagent}, which dynamically rendered news to readers, demonstrated that visualizations with dynamic designs of animated text could make information presentations entertaining.
Prior research~\cite{Forlizzi03Kinedit, Wang04communicate, Lee06Kinetic, yeo2008kim} has studied generating animated text automatically, assuming that simple animation schemes are good enough for an enjoyable experience and can save design effort.
For instance, Minakuchi and Tanaka~\cite{Minakuchi05Kinetic} designed an automatic typography composer that allowed semantic matching between keywords and motion schemes.
In general, past investigations worked on a limited set of animations for distinct emotions. However, our approach adjusts some low-level configurations in a given scheme to dynamically control the intensity of the expressed emotion.

On the one hand, the myriad techniques for single animated text~\cite{malik09communicating} are not readily applicable for designing animated wordle because the integrated effects of multiple text elements require special considerations~\cite{Tversky02facilitate}.
For example, a single bouncing ``Hello'' can be vibrant, yet twenty jumping greetings might be annoying.
On the other hand, a few techniques for animating text documents were not intended for communication purposes.
For instance, Fluid Document~\cite{chang1998negotiation} explored novel browsing experiences by revealing web page content on demand.
Diffamation~\cite{chevalier2010using} supported comparing two document versions through text animation.
There is little study of the animated text in the data visualization literature, despite a rich palette associated with text elements~\cite{brath20VisWithText}. 
Our work contributes an initial effort in designing animated wordles for emotion expression.

\subsection{Affective Visualization Design}

Expressing emotion has been widely studied in the human-computer interaction community~\cite{van12design, ma2021review}, covering design elements including pattern~\cite{urquhart17emotive}, texture~\cite{felecia15textility}, typeface~\cite{odonovan14font}, font style~\cite{Kalra05TextTone}, speech balloon shape~\cite{aoki2022emoballon}, and generative art~\cite{krcadinac2015textual}.
Recent years have witnessed increasing scholarly attention to the emotional dimension of data visualization~\cite{wang19value, lee2022affective}.
Feng \etal{}~\cite{Feng17MotionScape} termed research in the use of visual elements to change the affective nature of a visual representation as ``affective visualization''.

Empirical studies (\eg, \cite{kennedy18feeling, Lan21Smile, Lan2022Negative, lan2022chart}) summarized general patterns or guidelines for visualization design from real-world instances.
Besides, some work concentrated on specific design factors.
Bartram \etal{}~\cite{Bartram17color} presented a discrete color palette for eight distinct emotions, which was derived from image mining and crowdsourcing ratings.
Based on their work, Kulahcioglu \etal{}~\cite{Kulahcioglu20affect} further studied the emotive aspect of font families and supported creating emotionally congruent word clouds.
Anderson and Robinson~\cite{anderson2021AffectMap} compared the influence of applying affectively congruent and incongruent color palettes on map reading.

Among various design factors, this study looks into animation in particular.
It has been long recognized that animation is beneficial for visualization presentation~\cite{fisher10animation}.
Chevalier~\etal{}~\cite{Chevalier16animation} further pointed out the role of animation in conveying emotion. 
However, animating visualization designs for communicating emotion is still largely underexplored. 
Bartram and Nakatani~\cite{Bartram2009Parameters} explored how attributes of expressive motions, \eg, velocity, fluidity, path shape, \etcc{}~influence emotion conveyance.
They obtained experiment stimuli from human performers' gestures; therefore, the result was difficult to generalize to visualization designs directly.
Feng \etal{}~\cite{Feng17MotionScape} explored compositions of visual forms in motion to express emotions in immersive environments.
\draft{Most relevant, Lan \etal{}~\cite{lan21kineticharts} studied 60 cases of affective animated visualizations and summarized 20 design patterns.
However, their scope did not cover wordle, and the cases served as references without being implemented into authoring templates.}
Focusing on wordle, our work takes an initial step to democratize affective visualization creation, where we investigated parameter-controlled animation design to incorporate various user-input data.



\subsection{Wordle Variants}
Wordle is effective for text summary and topic understanding~\cite{tag2007rivadeneira,felix17taking}.
However, as a wordle maps word frequency to its size, people's perception can be easily influenced by the word size, which may hinder analytical tasks.
To address this issue, research in wordle design has focused more on supporting fast authoring (\eg{},~\cite{Koh10ManiWordle,Jo15WordlePlus,Wang18EdiWordle}) for communication or complementing semantic or temporal information of the content.
Our work shares the same goal to aid flexible creation by extending wordle's visual encoding and suggesting extra information.

To encode extra information, most studies took advantage of the natural visual channels of wordles.
For instance, the neighboring regions were used to indicate semantically relevant words~\cite{wu11semantic, xu16semantic, hearst2020semantic}, while the wordle contour also implies the theme of the content~\cite{chi15morphable,wang20shapewordle}. 
DancingWord~\cite{shu20dancingword} leveraged the movement of wordle to imply a change of scene in storytelling~\cite{shu20dancingword}.
Sparklines~\cite{Lee10SparkClouds} integrated external visualization with words.
WordStream~\cite{Dang19WordStream} embeds words in another visualization form, like a theme river.
Despite much exploration, few works consider the emotional context of wordle.
Most relevant, Kulahcioglu \etal{}~\cite{Kulahcioglu20affect} explored the affective effect of fonts and presented an interface for recommending affect-congruent wordle in terms of a scheme of font and color.
In comparison, we investigated the understudied aspect of animation in wordle~\cite{brath20VisWithText,Hicke2022wild}.
Our approach has no constraints on the wordle layout and can thus be integrated with many existing works.

\section{Design Goal}
\label{sec:goal}
Emordle is a new concept referring to a class of animated word clouds that deliver the emotional context underlying a wordle.
We believe it is beneficial for data communication.
However, it is unknown yet how to animate wordles to convey emotions.
To direct the exploration, we reflected on the design goals of an \abbr{} generation method.
We based this on our observation of wordle tools and previous experiences of visualization authoring tool research.

Wordles are found to be popular among the public due to the participatory culture, where people can engage in the creative authoring experience~\cite{viegas09participatory}.
We anticipated the \abbr{} creation process to bear a similarity, allowing people to superimpose an emotional response through the animation of wordles with ease.
It falls within the category of casual visualization~\cite{pousman2007casual}, or personalized visualization (\eg{},~\cite{Kim2019DataSelfie,zhang2020DataQuilt}).
As such, we summarized three design goals for \abbr{} generation.

\textbf{G1. Support the delivery of the intended emotion.} The method should support the \abbr{} creator to generate a credible result that reflects the intended emotion faithfully. Note that we assume the \abbr{} creator is very familiar with the underlying emotion.

\textbf{G2. Facilitate fast generation}~\cite{wordArt,Jo15WordlePlus}. Wordle data contains text content and its weight. The generation method should be able to provide an animated wordle in congruence with the given emotion within a few human interactions. The configuration should be minimal to reduce the manipulation cost and the degree of uncertainty among configuring factors. With the real-time generation of animated wordle, users \rev{may} iterate quickly on the animated schemes and derive a satisfying one. 

\textbf{G3. Enable personalized visual appearance}~\cite{Koh10ManiWordle, Wang18EdiWordle, Kim2019DataSelfie}. As audiences are sensitive about unconventional things~\cite{ortony1990cognitive}, the method should be flexible enough to harness the authors' creativity, support creating diverse final outputs, and thereby encourage the participatory design~\cite{viegas09participatory}.


\section{Analysis of Animated Text}
\label{sec:design-space}

We conducted qualitative analysis on a self-curated corpus of online examples to gain deeper insights into designing animated wordles that express abstract emotions (G1).
Due to the limited instances of animated wordles being found, we expanded the selection criteria to animated text and collected $77$ clips, where $57$ had identifiable emotional designs.
We posted two questions to guide our analysis: (i) what are the building blocks of an animated text, and (ii) how to make text animations emotional?
\begin{figure*}[t]
\setlength{\abovecaptionskip}{5pt} \setlength{\belowcaptionskip}{-10pt}
\centering 
  \includegraphics[width=\textwidth]{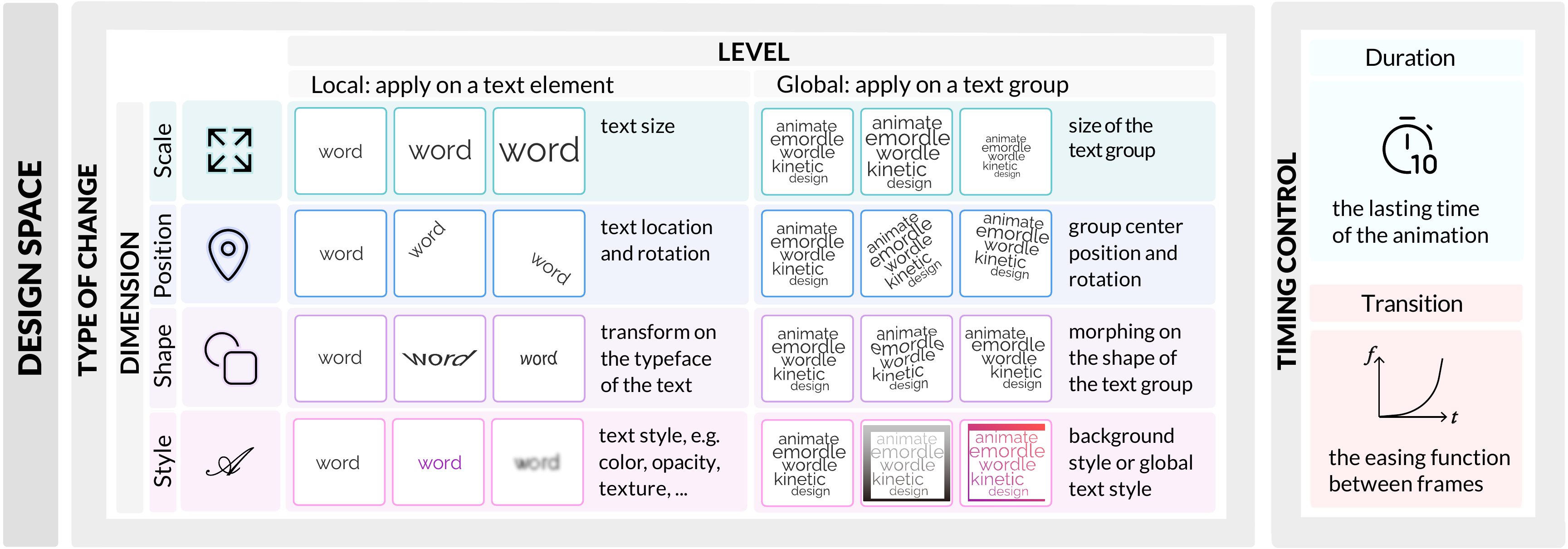}%
  \caption{A design space of animated text(s). The top layer compasses the type of change and the timing control of change. When multiple text elements appear, the animation types can be further decomposed into the local and global levels on top of the four dimensions of the visual design. The timing control includes the duration setting and transition design. We followed this design space when designing the animation scheme of emordle cases.}
  \label{fig:space}
\end{figure*}

\subsection{Collection} 
The search started with query terms including ``animated text'', ``kinetic typography'', and ``animated word cloud'' on Google.
The initial collection originated from sources including tutorials about animation tricks, collections of kinetic typography designs, miscellaneous clips from lyric videos, advertisements, and video openings, where text animation is heavily adopted.
We also included two demo videos from the literature~\cite{Lee02KineticTypographyEngine, yeo2008kim}.

Since most software decomposes animation into keyframes and timelines, we followed this paradigm and initially marked the changing properties of the graphical text elements, such as shape and size.
The categories of these properties gradually evolved as we discovered new examples.
The process ended when no more \rev{categories were} identified in the consequent two pages of the search engine results.
We did not take in new examples with an animation scheme similar to existing ones, such as transforming text into icons.
This was because we aimed to cover different animation schemes at a high level rather than constructing a comprehensive dataset.


\subsection{Coding}
Two of the authors coded the examples together, viewing them individually. The material is available at \survey{}.
For each entry, we labeled (a) the animation scheme details and (b) the perceived emotion(s). In addition, we used (c) free text to note what the animation was like and our associations to real-life scenarios based on the animated scheme.
As emotions \revv{are} naturally ambiguous and might be subjective across the audience, we conducted the coding to gain an initial probe of understanding instead of a strict result of taxonomy.

Guided by how the major graphics authoring tools (\eg{}, After Effects~\cite{aftereffects}) decompose an animation, we chose to code the animation scheme (a) based on the animation target, dynamic text properties, and timeline design (speed and transition types between keyframes).
The animations in the collection exhibit dynamic properties in terms of scale, position, shape, and style.
The speed was divided into three levels: fast, medium, and slow.
For convenience, we categorized transitions into linear, bump, gradual easing (slow in, slow out, slow in-out, and fast in-out), or mixed.
The former three are common transition types in keyframe-based animation designs, and they are easily distinguished.

The perceived emotions (b) were labeled using terms in the Geneva emotion wheel (version 3.0, abbreviated as GEW model)~\cite{scherer13grid}, which is a common measurement instrument for self-reporting an emotional experience.
By turn, we communicated which emotion was recognized and explained which design aspects led to the impression.
We also agreed on an appropriate emotion class and intensity for the most effective clip in delivering emotions. 
When we had different opinions about an emotion, such as no emotion versus joy, we tried to reach an agreement or abandoned it.
We excluded $20$ examples that barely elicited any emotion.
These \revv{are} mostly from advertisements or brand designs, where we found the animation largely served to draw attention.

\subsection{Design Space of Animated Text}
By decomposing animation schemes (a) of the examples in the corpus, we derived a design space for the low-level building blocks of animated text (i).
The space has two layers.
One is concerned with the type of change, and the other corresponds to the timing control of change.

The type-of-change layer can be further divided into two levels.
The \textbf{local level} (66\%) applies to the atom elements of words, while the \textbf{global level} (23\%) applies an animation scheme to a group of words globally.
Our analysis on the global level was based on $17$ examples whose animation target included more than five text elements, which we deemed visually similar to wordles.
Most cases we found \revv{use} all text elements as a group.
Words \revv{are} also grouped according to sentence co-occurrence or layout neighborhood.
Under each level, there are four major dimensions for animation design: scale, position, shape, and style (see~\autoref{fig:space}).
We found that the local-level animation design \revv{could} be combined with the global to a certain degree in most cases.
Below we explicitly describe these dimensions and report their appearance ratio among the 77 clips.

\begin{enumerate}[nosep, leftmargin=15px]   
\item[\inlinegraphics{icons/scale.pdf}]  The \textbf{\textit{scale}} (17\%) refers to the size of the word or word group respectively. For instance, sudden inflation might highlight a high-frequency word with exaggeration, and a global shrink might mimic a shrug.

\item[\inlinegraphics{icons/location.pdf}] The \textbf{\textit{position}} (66\%) is relevant to the location and rotation of the word anchor locally or the group center globally. While each text element can have a small moving offset, such movement can be coordinated by its layout position from a global aspect.

\item[\inlinegraphics{icons/shape.pdf}] The \textbf{\textit{shape}} (31\%) of a local text element refers to its transformation from its original typeface. One example is the calligram (\eg{} ~\cite{Zou16calligram, Maharik11Micrography}) that distorts a text according to the vector fields of a shape constraint. The other emerging example is the variable font, whose typeface properties (\eg, italic extent, font weight, and character width) can be controlled by continuous parameters. The global \textbf{\textit{shape}} refers to the overall shape of the word group (\eg{} ~\cite{chi15morphable, wang20shapewordle}).

\item[\inlinegraphics{icons/style.pdf}] The \textbf{\textit{style}} (30\%) dimension comprises numerical styling properties on graphical elements. Common local \textbf{\textit{style}} include color attributes like hue and saturation, opacity, texture, and blur. Note that discrete text attributes (see Brath's review~\cite{brath20VisWithText}) like typeface and case do not belong to this class, as we focus on the continuous type of change. The global \textbf{\textit{style}} is similar, despite the effect being based more on the global aspect, such as a dynamic color gradient being applied to all texts.
\end{enumerate}

\vspace{\baselineskip}

The other layer concerns the speed of change that accompanies animation.
An even more subtle narration can be achieved by properly controlling the timing. The \textbf{\textit{duration}}  \inlinegraphics{icons/timer.pdf} is the basic configuration for an animation that determines its lasting time. Another fine-grain control is the \textbf{\textit{transition}} \inlinegraphics{icons/transition.pdf} that arranges the duration for each of the changes to the graphical elements. In practice, animators produce keyframes to anchor the state of the transition and leverage a simple easing function to play with the progress of time instead of transiting states smoothly. Apart from using only one transition type, 36\% of examples \rev{have} adopted a mixed timeline design for staging.

\subsection{Design Emotions in Text Animations}
According to our corpus coding, the detailed distribution of the 64 tagged emotions was as follows: amusement--6.3\%, joy--6.3\%, pleasure--12.5\%, tenderness--12.5\%, wonderment--12.5\%, surprise--7.81\%, relief--4.7\%, longing--4.7\%, fear--9.4\%, despair--7.8\%, repulsion--7.8\%, and anger--7.8\%.
Note that 4 examples had two tags.
By aligning the emotion tags (b) with the description of the animation design (c), we analyzed the high-level strategies (denoted as \textbf{S}) for crafting emotions in text animation schemes (ii). 

\textbf{S1. Simulate real-life scenarios}.
Sixty percent of emotional animated text instances \revv{adhere} to the basic laws of physics.
They \revv{apply} classic techniques~\cite{thomas1995illusion} like ``squash and stretch'' (maintaining object volume during the deformation) and ``slow in/out'' (taking the acceleration speed into consideration).
Such a setting contributes compelling visual effects, surpassing the clumsy linear interpolation between keyframes.
Several instances reminded the coders of natural phenomena associated with personal feelings, like a stormy wave for anger and breezy wind for tenderness.
We also noticed a special interest in particle effects that break the character into finer pieces for even smoother morphing into other shapes.
Additionally, some works \revv{mimick} physiological phenomena.
For instance, a shaky word \revv{mimick} trembling out of fear;
and a gradual blurring effect \revv{is} like someone being lost in reverie.

\textbf{S2. Portray linguistic features}.
As suggested by Forlizzi \etal~\cite{Forlizzi03Kinedit}, portraying linguistic features of the text content \revv{helps} deliver emotions since word semantics are naturally presented with the text element.
In our corpus, there \revv{are} 23\% works that adopt such a strategy.
For example, words' going upwards \revv{looks} like a rise in pitch, indicating a state of surprise.
Increasing the font weight \revv{is} analogous to raising the volume of one's voice, where the same \revv{applies} to a sudden increase in the text size.

\textbf{S3. Coordinate the randomness and speed}.
For examples involving multiple text elements, the randomness and the speed of individual elements \revv{plays} an essential role in delivering emotions.
This \revv{echoes} prior findings on particle systems where multiple graphical elements were involved (\eg{},~\cite{Feng14MotionScape}).
For those labeled as ``repulsion'', they \revv{share} a common feature: unpredictable changes, such as a glittering background color, twinning together and rotating in 3d, and so on.
There \revv{are} two animations that simply mimic the waves yet \revv{are} tagged as different emotions.
One \revv{is} like a calm, slow ripple, whilst the other \revv{moves} like a storm surge at fast speed.
In summary, the rhythmic, ordered, or at least predictable animations easily \revv{create} an aesthetically pleasing experience.
On the contrary, unordered movements \revv{add} fuzziness, which \rev{could} arouse negative feelings.

\section{Creating an Emordle}
\label{sec:method}

With the high-level strategies to inject emotions and low-level decomposition of animated schemes identified through the corpus analysis, we continued to explore a feasible (G1) and novice-friendly (G2) solution for \abbr{} generation applying the insights.
In this section, we introduce a composite approach that enables end-users to author an \abbr{} effortlessly.
Guided by the approach, we designed \abbr{} cases for four basic emotions.

\begin{figure*}
\centering
  \includegraphics[width=\textwidth]{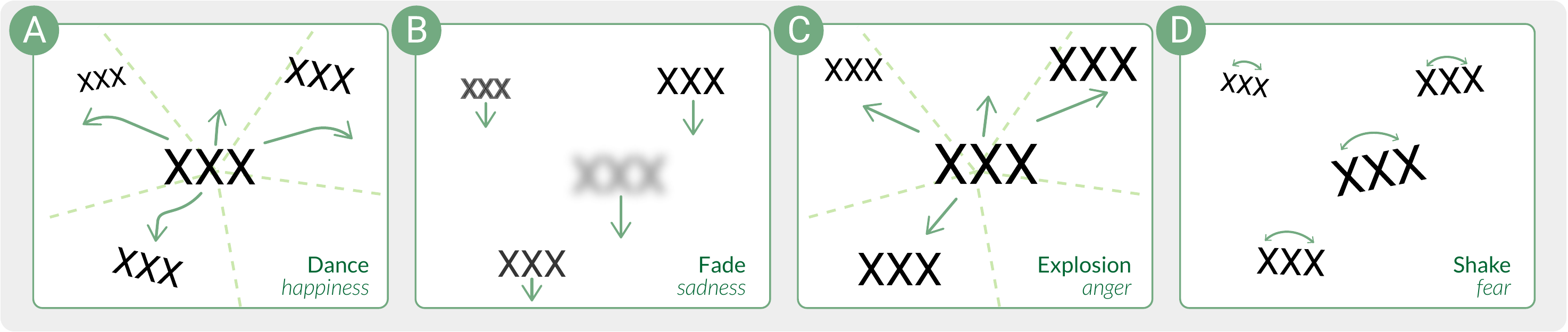}
  \caption{An illustration of the four animation schemes in our study. (A) \textsc{Dance} (\textit{happiness}): words spread out and slightly rotate. (B) \textsc{Fade} (\textit{sadness}): texts turn to a deeper color, drop down, and become blurred. (C) \textsc{Explosion} (\textit{anger}): words grow large and fast to the corners. (D) \textsc{Shiver} (\textit{fear}): words tremble and randomly move around.}
  \label{fig:case}
\end{figure*}

\subsection{A Composite Approach for Emordle Generation}
We proposed an extensible approach to construct an \abbr{}.
In general, an \abbr{} is created by a pre-defined animated word design that determines the emotion class and is later fine-tuned to fit the emotion intensity through two parameters: entropy and speed.

Our approach assumes that there is one and only one intended emotion in the associated wordle data, \ie{}, there is no conflict with the underlying emotion of each word in their intended semantic.
Such occasions might happen when the wordle summarizes a narration, like an individual's story with both ups and downs.
However, there are hardly any guidelines to attain expressiveness and affective capacity in an animation design, as the human perception of emotion relates to the interplay of multiple factors.
Prior research \rev{has often studied} one particular design aspect only and \rev{controlled} others (\eg{},~\cite{desai19geppetto}).
In addition, the goal is to produce animation schemes emotionally congruent with the wordle context.
We admit the dominant influence of affective response in the language itself, holding the common belief that the animation can augment the emotion underlying the original context, but not reverse the original emotion~\cite{Lee02KineticTypographyEngine}.

\textbf{Reusing Animated Text Design.}
Treating a wordle as a combination of words, an animated wordle could be constructed by assigning different animation schemes for each word.
It is established that an animated text can deliver emotion~\cite{Lee02KineticTypographyEngine}.
We further hypothesize that applying a given emotional animation scheme for all words in a wordle yield a similar impression of emotion (\textbf{H1}).
The design of an emotional animation scheme for a word is left to experienced designers, which is a simpler task than arranging multiple words.
Following the corpus analysis, we used keyframe-based animation to break up animation into programmable units converted from the original design. It is a basic paradigm in animation authoring and achieves fair expressiveness.
Keyframes of a wordle are used to manipulate the type of change to each word, including position \inlinegraphic{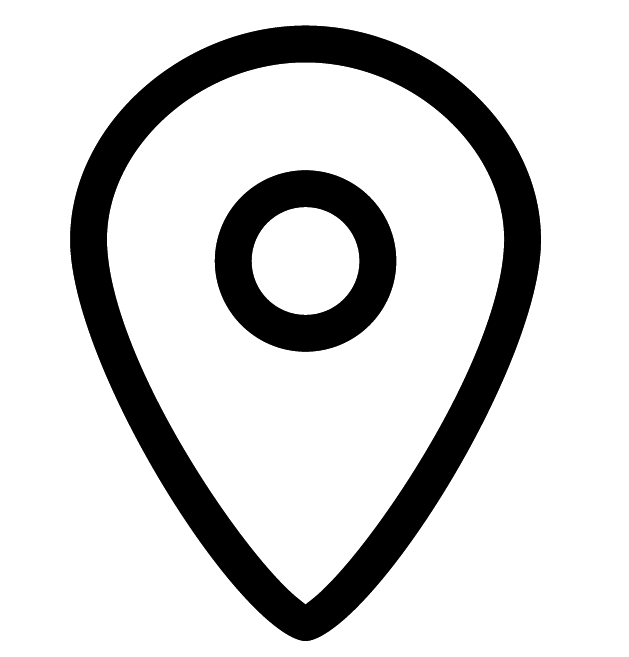}, rotation \inlinegraphic{figures/icons/location.pdf}, size \inlinegraphic{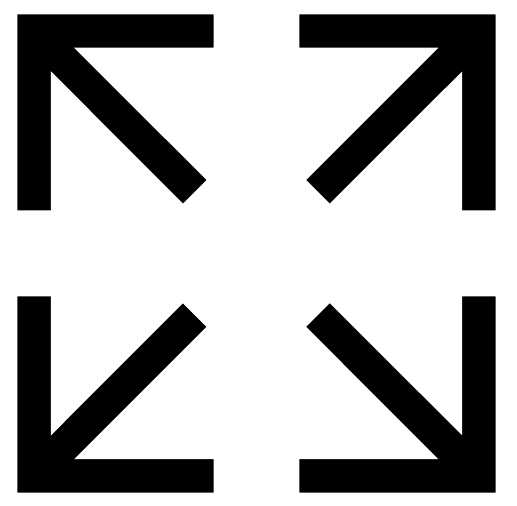}, opacity \inlinegraphic{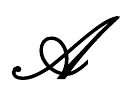}, color transforms \inlinegraphic{figures/icons/style.pdf}, and font attributes \inlinegraphic{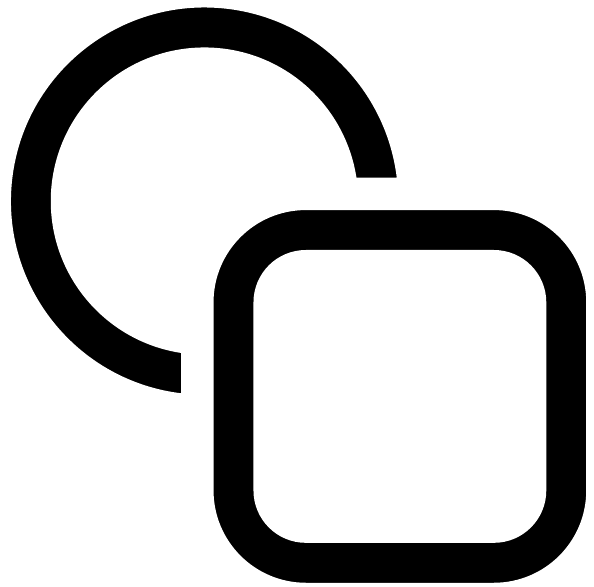} (as in variable fonts).
Transitions \inlinegraphic{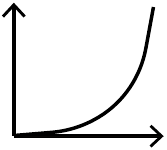} between two neighboring frames are determined by a suite of easing functions interpolating them.


\textbf{Coordinating Text Group Animation with Entropy \& Speed.}
We propose to adjust its low-level animation properties according to entropy and speed, as informed by the design patterns in the emotional animated text (S3).
Specifically, we propose the following strategies to incorporate these two dimensions as tunable parameters into the animation scheme design.

Entropy governs the degree to which the animation exhibits chaoticness, that is, randomness.
Inspired by examples in the corpus, entropy can be controlled by the number of word groups and the extent of their movement.
There are multiple ways to distribute words into groups.
For instance, they can be divided according to the positional adjacency, along the horizontal/vertical direction or in the 2D space (applying the k-nearest-neighbors algorithm).
Otherwise, they can be randomly assigned.
Each group can be assigned animation schemes with variations to the backbone animation.
The scale \inlinegraphic{figures/icons/scale.pdf}, position \inlinegraphic{figures/icons/location.pdf}, shape \inlinegraphic{figures/icons/shape.pdf}, and style \inlinegraphic{figures/icons/style.pdf} of words are recalculated based on the allocated keyframe configuration.
The specific animation scheme can vary slightly inside individual groups, thereby contributing to the global-level type of change, \eg, moving direction \inlinegraphic{figures/icons/location.pdf} and the delay of the animation \inlinegraphic{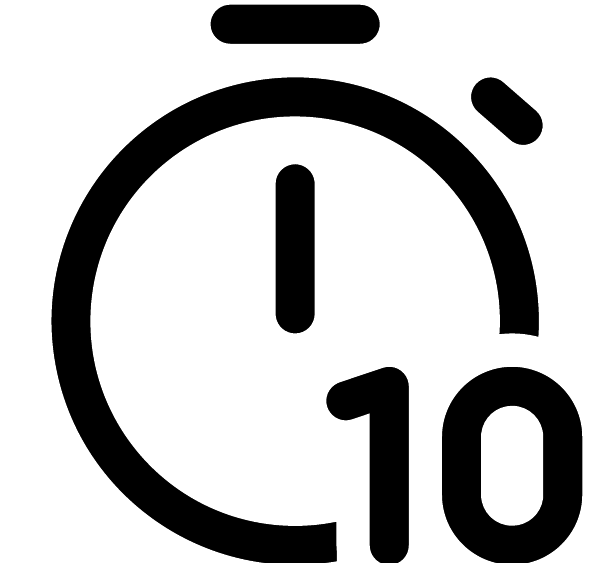}.
In this way, an increasing group number makes the overall animation manifest a greater degree of randomness.
In extreme cases, each word belongs to its own group.
On the other hand, higher entropy also exaggerates the animation by amplifying the extent of the reach of the word groups.
To some extent, this approach alleviates the design burden of tweaking the low-level animation properties of every word. Thus the animation is controlled holistically and only a small part of the existing animation scheme shall be modified.

Another parameter, speed, is familiar to the public as it only concerns the duration of one animation loop.
Internally, we allocate a different number of keyframes according to the value of the speed.
Higher speed squeezes additional keyframes into the same time duration, which causes the wordle's movement to be swifter.

\subsection{Design Emordle Cases}
\label{sec:design}
Guided by the composite approach aforementioned, we developed four design cases, each for a basic emotion category (see~\autoref{fig:case}).
These four cases demonstrate how to instantiate the composite framework for different emotion classes and are adopted in the later controlled experiments as concrete stimuli to test our hypotheses. 


The backbone animation schemes for individual texts were reproduced from compelling examples in the corpus.
Specifically, \textsc{Dance} implies \textit{happiness} with a dancing effect (S1), where words spread out and slightly rotate.
\textsc{Fade} is analogous to tears falling (S1), which suggests \textit{sadness}.
Inside each word group, texts would turn a darker hue, drop down with ease, and become blurred, and then gradually fade away.
\textsc{Explosion} uses an explosion metaphor to imply \textit{anger}, where words pop out with an elastic transition on an increasing scale as if being shouted out (S2).
\textsc{Shiver} makes words shiver (S1) in correspondence to the \textit{fear} emotion.
Words tremble left and right quickly while dropping down.
For coordinating the animation of multiple texts, we followed a grouping strategy and designed variance in the group-specific animation as detailed in~\autoref{tab:case}.
\begin{table}[htbp]
\caption{Group animation schemes of the \abbr{} cases.}
\label{tab:case}
\centering
\tabcolsep=0.1cm
\begin{tabular}{lll}
\toprule
\textbf{scheme}  & \textbf{grouping} & \textbf{group-specific parameter}  \\
\midrule
\textsc{dance} &   k-neighbor  & moving direction/distance \& rotation angle \\ 
\textsc{fade}  & random & delay \\
\textsc{explosion} & k-neighbor & delay, moving direction/distance \\
\textsc{shiver}  & random  & delay, rotation angle \& dropping distance\\
\bottomrule
\end{tabular}
\end{table}

\begin{figure*}[ht]
\centering
  \includegraphics[width=\linewidth]{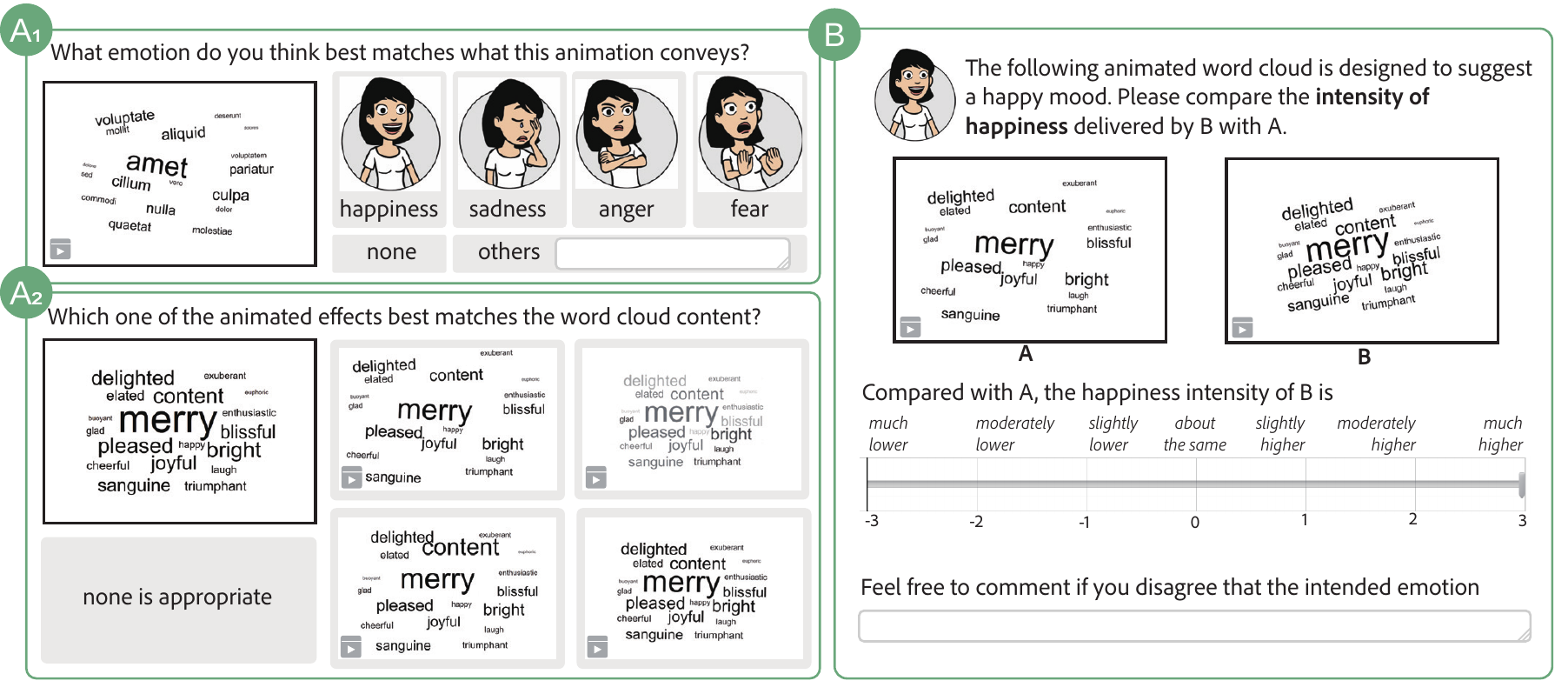}
  \caption{An illustration of the crowdsourced user experiments. (A\textsubscript{1}) Study I.1: Choose the conveyed emotion from an \abbr{} without semantics. (A\textsubscript{2}) Study I.2: Choose the most suitable animation scheme that matches the context. (B) Study II: Compare the intensity of the expressed emotion.}
  \label{fig:crowd}
  \vspace*{-\baselineskip}
\end{figure*}

During the design process, we continually consulted one designer with three years of experience in authoring data videos to refine the animation schemes, especially for group-specific variances.
The animation scheme refinement focused on the following aspects.

(1) \textit{Maintaining readability.} Abrupt changes in text elements might lead to readability issues.
There is a trade-off between exaggerating the animation for emotional impact and conveying textual information to the audience. The more exaggerated the animation, the more likely it is to imply emotions, but it may also lead to a decrease in the audience's ability to comprehend textual information. Therefore, we empirically determined the maximum speed and entropy of the animation to ensure that it does not interfere with the audience's reading experience.

(2) \textit{Preserving parameter consistency.} Each animation scheme has a relatively fixed pattern of keyframe settings, like what type of change should be included.
The exact form or extent of change is dependent on the detailed keyframe entry and state values.
To tune the emotion expression, we imposed two high-level parameters, \ie{}, speed and entropy, for keyframe values on top of these schemes according to our animation decomposition strategy.
However, their relationships with the minute features of the animation scheme remain open, especially for entropy, which is even more abstract.
We aligned the keyframe state to parameters by first setting up the extreme text property and then preserving the change in a consistent way.

\subsection{Build Up an Emordle Generation Tool}
The composite approach takes in a pre-set animation scheme and produces modified animated schemes with different emotional intensities coordinated by speed and entropy.
It conforms to the GEW model~\cite{scherer13grid} employed in our user study, a psychological measure of emotions for self-report, representing an emotion with basic emotion classes (N=20) and corresponding intensity.

To support end-users under a casual context, we envision an open design repository with a fair amount of \abbr{} cases that cover all basic emotions (G1) with varieties (G3).
For instance, 20 basic emotions with three variations each.
With the repository, end-users may create an emotionally congruent \abbr{} by choosing from various animation schemes under a specific emotion class and tuning the speed or/and entropy to arrive at the desired level of emotion intensity.
These operations can be done on a simple user interface with buttons and sliders, which meets our anticipation in supporting fast authoring (G2).
While it is possible to pre-define a set of animation schemes and directly generate \abbr{}s matching the declared emotion, we keep humans in the creation process.
This is because different wordle data may influence the delivery of emotion and hence require validation; people may also need to explore potential effects.

Meanwhile, designers or enthusiasts may contribute to the repository by creating diversified backbone animated schemes associated with one particular basic emotion.
They may refer to the identified strategies for designing emotions in text animations and follow a similar path as how we produced the design cases.
Note that how to design a backbone animation scheme effectively is beyond our scope.

\section{Crowdsourced User Experiments}
\label{sec:eval}

We conducted two crowdsourced user experiments to evaluate whether the composite framework yielded the emotional wordle animation as intended. Using the four design cases as stimuli, the first study initially confirmed the crowd agreement on emotion over specific animation schemes (\textbf{H1}), and the second one demonstrated that speed and entropy affected the expressed emotion intensity (\textbf{H2}).

\subsection{Set-up}
Both studies were performed on Prolific\footnote{\url{https://app.prolific.co/}} based on Qualtrics\footnote{\url{https://www.qualtrics.com/}} questionnaires.
We ran a pilot study with 50 participants to decide on questionnaire details and the sample size.
The participant pool was prescreened with the following requirements to ensure the response quality: (i) fluent in English, (ii) $100\%$ acceptance rate, and (iii) at least $20$ rounds of crowdsourcing experiences.
We distributed the study to a balanced sample with an even number of male and female participants.
Each participant \rev{was} paid £$7.2$ hourly.
In general, the questionnaire started with an introduction to wordle and asked about participants' knowledge of wordle.
Then participants were shown \abbr{} gifs and asked to evaluate the delivered emotion through single-choice questions or free text.

\textbf{Word List Synthesis.} 
We synthesized animated wordles as stimuli based on five word lists. Each wordle had $18$ words, which is on a moderate scale in online instances of our corpus.
One word list, named \texttt{lorem}, included placeholder texts~\cite{lorem} without any semantics.
The other four word lists were synonyms of the four emotion classes in the design cases, namely \texttt{happiness}, \texttt{sadness}, \texttt{anger}, and \texttt{fear}.
To avoid suggestive semantics that influence participants' judgment, we elaborately removed any relevant words implying animation details, such as ``swing''.
 
\textbf{Static Wordle Configuration.} 
Considering a large design space of animated wordles with various dimensions,
we made the following constraints on factors that influence emotion delivery in an \abbr{}.
(1) \textit{Maintain a consistent layout across stimuli}. To maintain the layout consistency across different study settings, we sorted the words of the individual dataset according to the word length.
We then assigned the font size and position of words with the same rank number derived from a randomly generated number. 
While some wordles \revv{leverage} particular shapes (\eg{}, \cite{chi15morphable, wang20shapewordle}) to convey extra information, we adopted the rectangular layout in the controlled study to reduce the impact of any relevant factors. 
All stimuli \rev{shared} the same canvas size and shape.
(2) \textit{Apply the same typeface and color set for text elements}. Previous work has demonstrated the usage of color palettes~\cite{Bartram17color} and typefaces~\cite{Kulahcioglu20affect} to deliver emotions.
To control the emotive aspect of the text style, we adopted a black hue and a regular typeface (Arial).
Note that these constraints \rev{were} designed for the controlled study to specifically evaluate animation schemes. 
Creators \revv{can} still combine these dimensions (\eg{} outline, color scheme, and typeface) in designing \abbr{}. 

\textbf{Animation Generation.} Given a static wordle configuration, we applied the animation schemes according to our design cases.
There was a $3 \times 3$ grid of parameters for speed and entropy for tuning each animation scheme.
The greatest speed led to the least readability, and the lowest speed was considered to be the least effective as determined by all authors.
As for the entropy, the largest value split words into 18 groups.
The lowest value formed only a few groups (1 for random split and 4 for positional split).
The middle parameters were the medium of the two extremes.

\subsection{Study I: Agreement on Animation Scheme}
Study I aimed to test our assumption that some particular animated text design would yield a common understanding of emotional intent (\textbf{H1}).
\begin{table}[t]
\centering
\tabcolsep=0.17cm
\caption{Answer distribution in Study I.1 (omitting the ``others'' option). Participants chose the most appropriate emotion type for the \abbr{} based on the \texttt{lorem} dataset. The majority choice is highlighted in bold.}
\begin{tabular}{lcccccc}
\toprule
\textbf{scheme} & \textbf{\textit{happiness}}&\textbf{\textit{sadness}} & \textbf{\textit{anger}} & \textbf{\textit{fear}} & \textbf{none} & \textbf{total}  \\
\midrule
\textsc{dance}   & \textbf{47.1\% }  & 9.8\% & 9.8\% & 3.9\% & 2.0\%        & 51  \\ 
 \textsc{fade}    & 5.9\% & \textbf{47.1\%} & 2.0\% & 9.8\% & 15.7\%          &  51  \\
\textsc{explosion}  & \textbf{25.5\%} & 5.9\% & 21.6\%      & 17.6\% &  9.8\%         &  51\\
\textsc{shiver}    & 8.8\% & 15.8\% & 5.3\%   & \textbf{40.4\%} &   17.5\%       &  57\\
\bottomrule
\end{tabular}

\label{tab:exp1-lorem}
\end{table}

\textbf{Configuration}. We divided the questionnaire into two parts, evaluating people's judgment of the animated wordles, with and without semantics.
In Study I.1, participants assessed the animation organically without semantic suggestions (see~\autoref{fig:crowd} A\textsubscript{1}).
We fixed the dataset to be \texttt{lorem}, and showed one \abbr{} applying one of the four animation schemes under the medium speed and entropy for each question.
We then asked which emotion best matched what the animation had conveyed.
Participants could choose from six options: \textit{happiness}, \textit{sadness}, \textit{anger}, \textit{fear}, others, and no emotion.
We encouraged them to explain their thoughts with free text in a follow-up question.
To facilitate the self-report on emotions, the textual options were augmented with comic figures in the PrEmo emotion measurement instrument~\cite{desmet2003measuring}.
Adhering to the basic assumption that one and only one intended emotion underlies the \abbr{}, we adopted the four emotion synonym datasets in Study I.2.
For each question, we showed four \abbr{}s using different animation schemes under the medium speed and entropy, and asked participants to choose one animation that best matched the wordle content (see~\autoref{fig:crowd} A\textsubscript{2}).
They could opt out by selecting ``None is appropriate''.

\textbf{Questionnaire}. To prevent participants from enumerating emotion types, we randomly selected three out of four emotion types for the two parts and shuffled the order for each questionnaire.
Therefore, there were 6 tasks in total.
We also put a single-choice question about the static wordle subject before showing the GIFs in the second part.
If a worker \rev{failed} to understand the underlying topic, the submission would be deemed unqualified.

\textbf{Responses}. We recruited $100$ participants and removed $30$ entries failing the quality tests.
As the question assignment exhibited randomness, each question had an average of 52.5 responses (SD=2.1).
78.6\% of participants understood the visual encoding of wordles and 17.1\% had experience in creating wordles. 
They spent 352.8 seconds (SD=262.4) on average to complete the study.
There were no outliers lower than the 1.5 interquartile range below the first quartile.

\textbf{Results \& Analysis}.
As seen from \autoref{tab:exp1-lorem}, when presented with no semantic clues, people had different ideas about the conveyed emotion, yet a slight convergence of opinion existed.
For the dance, fade, and shiver scheme, at least 40\% of the crowd recognized its intended emotion, which was significantly higher than a random choice between four options.
Besides that, each agreement rate was 20\% more than the second major selection.
Meanwhile, for the explosion scheme, 25.5\% of the crowd agreed on \textit{happiness}, 21.6\% agreed on \textit{anger}, and 17.6\% chose \textit{fear}.
We \revv{speculate} that the slightly higher ratio of \textit{happiness} \revv{is} because people \rev{might} get a positive impression seeing the text elements burst quickly out to the boundaries.

Despite the ambiguity in pure animation schemes, we witnessed increases in the agreement rates over the intended emotions in Study I.2 with the given context (see \autoref{tab:exp1-emotion}).
The majority choices for each question were \texttt{happiness}---\textsc{dance} (57.7\%), \texttt{sadness}---\textsc{fade} (82.7\%), \texttt{anger}---\textsc{explosion} (50.0\%), and \texttt{fear}---\textsc{shiver} (57.7\%).
These percentages were significantly above values when the participants made random selections (\ie{}, 25\%, ignoring the ``none'' option), and no less than 50\% (a threshold used in a prior study~\cite{lan21kineticharts}).
As such, there \revv{exist} positive signals that our four custom animation schemes of medium speed and entropy \revv{can} effectively convey the intended emotion.

\begin{table}[t]
\centering
\tabcolsep=0.148cm
\caption{Answer distribution in Study I.2. When the emotional context \rev{was} given, no less than half of the participants agreed on our intended animation schemes. The majority choice is highlighted in bold.}
\begin{tabular}{lcccccc}
\toprule
\textbf{dataset} & \textsc{dance}& \textsc{fade} & \textsc{explosion} & \textsc{shiver} & \textbf{na} & \textbf{total}  \\
\midrule
\texttt{happiness}   & \textbf{57.7\% }  & 13.5\% & 15.4\% & 7.7\% & 5.8\%        & 52  \\ 
\texttt{sadness}  & 1.9\% & \textbf{82.7\%} & 1.9\% & 11.5\% & 1.9\%          &  52  \\
\texttt{anger}  & 13.0\% & 7.4\% & \textbf{50.0}\%      & 18.5\% &  11.1\%         &  54\\
\texttt{fear}   & 9.6\% & 3.8\% & 26.9\%   & \textbf{57.7\%} &   1.9\%       &  52\\
\bottomrule
\end{tabular}

\label{tab:exp1-emotion}
\vspace{-2pt}
\end{table}

We went through the free-text comments to learn more about participants' opinions.
The \textsc{dance} scheme reminded people of other high-arousal emotions~\cite{russell80circumplex} like surprise, excitement, and anxiety.
The \textsc{fade} scheme led to an impression of calmness and pain.
The \textsc{explosion} scheme received the most diverse responses, including disgust, sorrow, pain, and excitement.
People felt pain and anxiety in terms of the \textsc{shiver} scheme.
The miscellaneous interpretations \revv{reflect} the ambiguous nature of animations and \revv{implies} the importance of evaluating potential interpretations when designing the backbone text animation scheme.

\subsection{Study II: Rating Emotion Extent}
Study II studied whether speed and entropy led to different conceptions of emotion extent (\textbf{H2}).

\textbf{Configuration}. We bound the emotion synonym dataset to the animation scheme assumed to be relevant and asked participants to rate the emotion intensity of the nine \abbr{} variants of (speed, entropy) tuples.
Prior to the formal study, we tested an alternative experiment design where participants had to provide an absolute emotion intensity based on a Likert scale for a given \abbr{}.
However, it was found that the responses varied significantly, which might result from a lack of consensus over the emotion intensity measures among different people.
Thus, we adopted paired comparisons in the formal study, which was a common approach in the literature~\cite{desai19geppetto, Wu21LQ2}.
We used the \abbr{} with the medium speed and entropy as the comparison base and placed another \abbr{} variant side by side (see~\autoref{fig:crowd} B).
Participants should report the emotion intensity in the stimulus based on a 7-point Likert scale.
$-3$ refers to much lower; $0$ means ``about the same''; while $3$ refers to ``much higher''.
As the questionnaire had primed participants for a given emotion type, we allowed them to indicate disagreement with an external question.

\textbf{Questionnaire}. There were 17 comparison tasks in a questionnaire, including one test for quality control.
For each animation scheme, four comparison tasks (eight in total) were randomly selected and assigned to the participants.
In the test, two identical GIFs were used and responses other than 1/0/-1 would be filtered out.

\textbf{Responses}. There were $150$ participants and $18$ entries were removed for failing the quality test.
78.0\% of them understood the wordle encoding and 16.7\% had created wordles before.
Their average duration was 449.1 seconds (SD=192.3) and there were no entries of an extremely short duration.
Each comparison task received an average of $64$ responses (SD=3.0).

\textbf{Results \& Analysis}.
~\autoref{tab:exp2} provides an overview of Study II.
The heatmap of average ratings implies a subtle change in emotion intensity among different (speed, entropy) configurations.
The \textsc{dance}, \textsc{explosion}, and \textsc{shiver} scheme share a similar pattern: low speed and entropy largely decrease the emotion intensity, whereas high speed and entropy lead to increased intensity.
On the contrary, the \textsc{fade} scheme has an increased intensity for higher speed and entropy.
This result corresponds exactly with the valence-arousal emotion model~\cite{russell80circumplex}, where \texttt{sadness} (\textsc{fade}) has low arousal while \texttt{happiness} (\textsc{dance}), \texttt{anger} (\textsc{explosion}), and \texttt{fear} (\textsc{shiver}) have high arousal.

\begin{table}[t]
\centering
\tabcolsep=0.1505cm
\caption{Results for Study II. Each $3\times3$ heatmaps in the mean and std column show the mean values and standard deviance under different settings of speed and entropy, where the x-axis represents entropy and the y-axis stands for speed (left to right/top to bottom: lowest to highest value).  **: 99\% confidential interval.}
\begin{tabular}[t]{lllllll}
\toprule
\textbf{scheme} & \textbf{dataset} & \textbf{\textsc{K-W} test}  & \textbf{mean} & \textbf{}  & \textbf{std} &  \\
\midrule
 \textsc{dance}  & \texttt{happiness} & 149.1** & \raisebox{-.37\height}{\includegraphics[height=25pt]{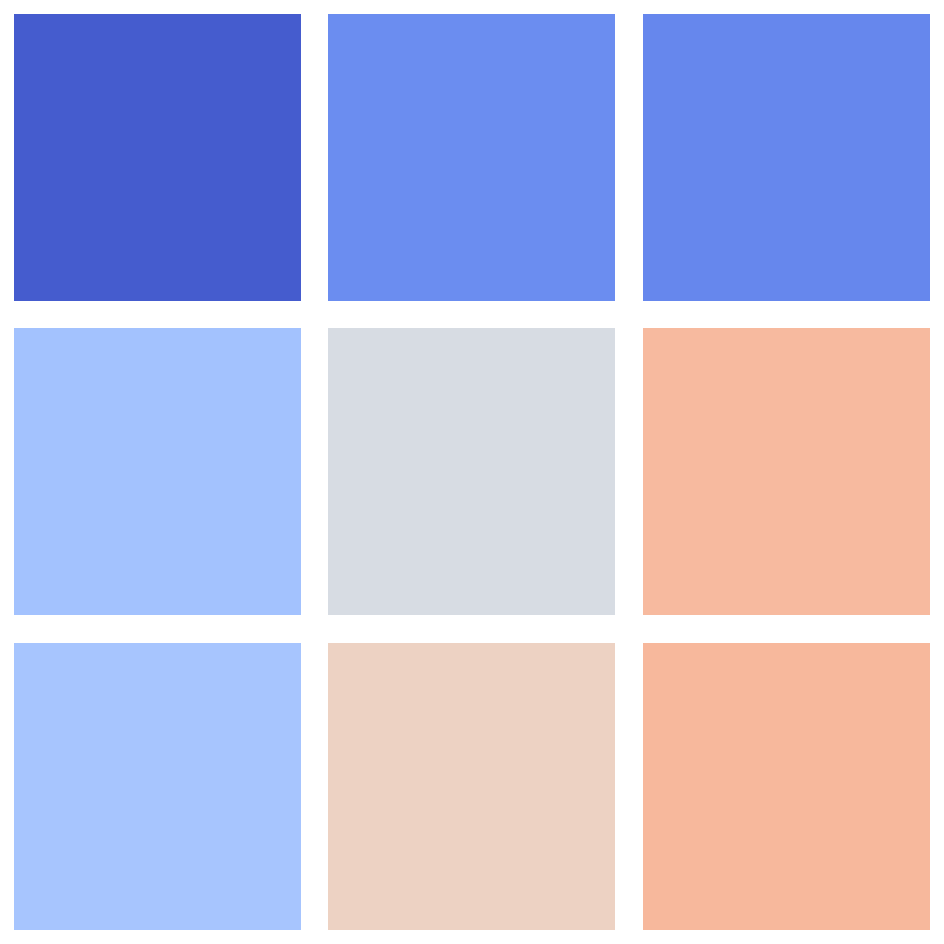}} & \raisebox{\height}{\multirow{4}{*}{\includegraphics[height=90pt]{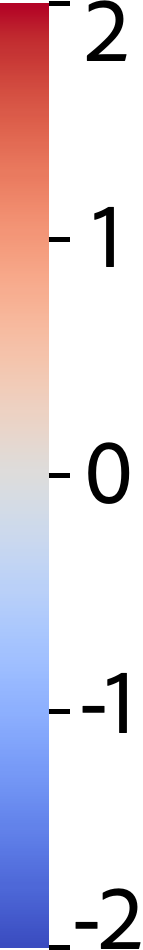}}} & \raisebox{-.37\height}{\includegraphics[height=25pt]{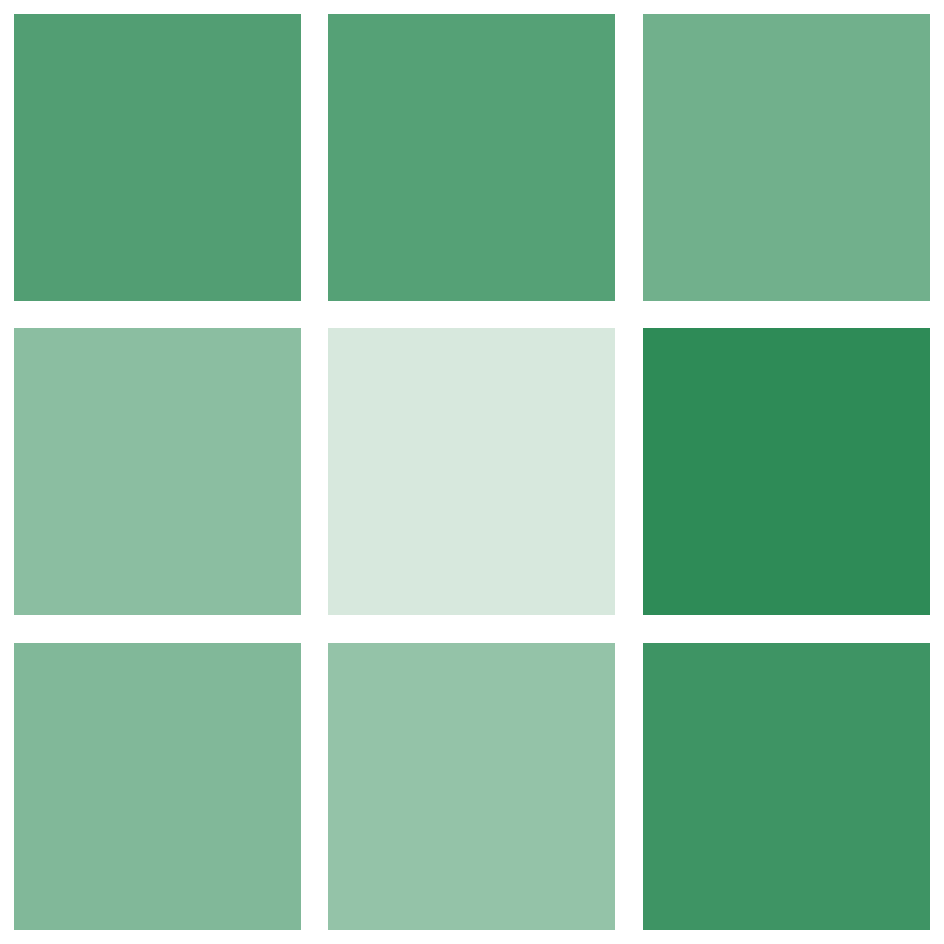}} & \multirow{4}{*}{\includegraphics[height=90pt]{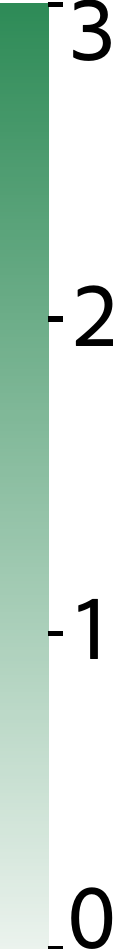}}\\[10pt] 
 \textsc{fade} & \texttt{sadness} & 31.3** & \raisebox{-.37\height}{\includegraphics[height=25pt]{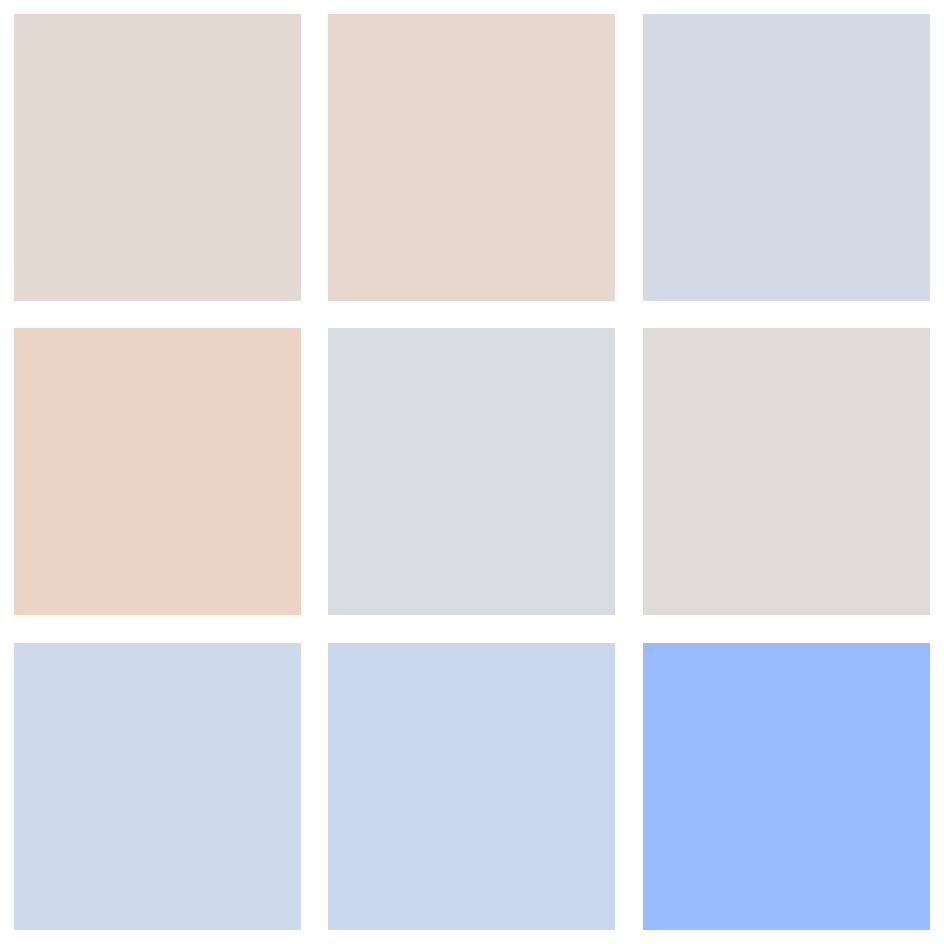}} & & \raisebox{-.37\height}{\includegraphics[height=25pt]{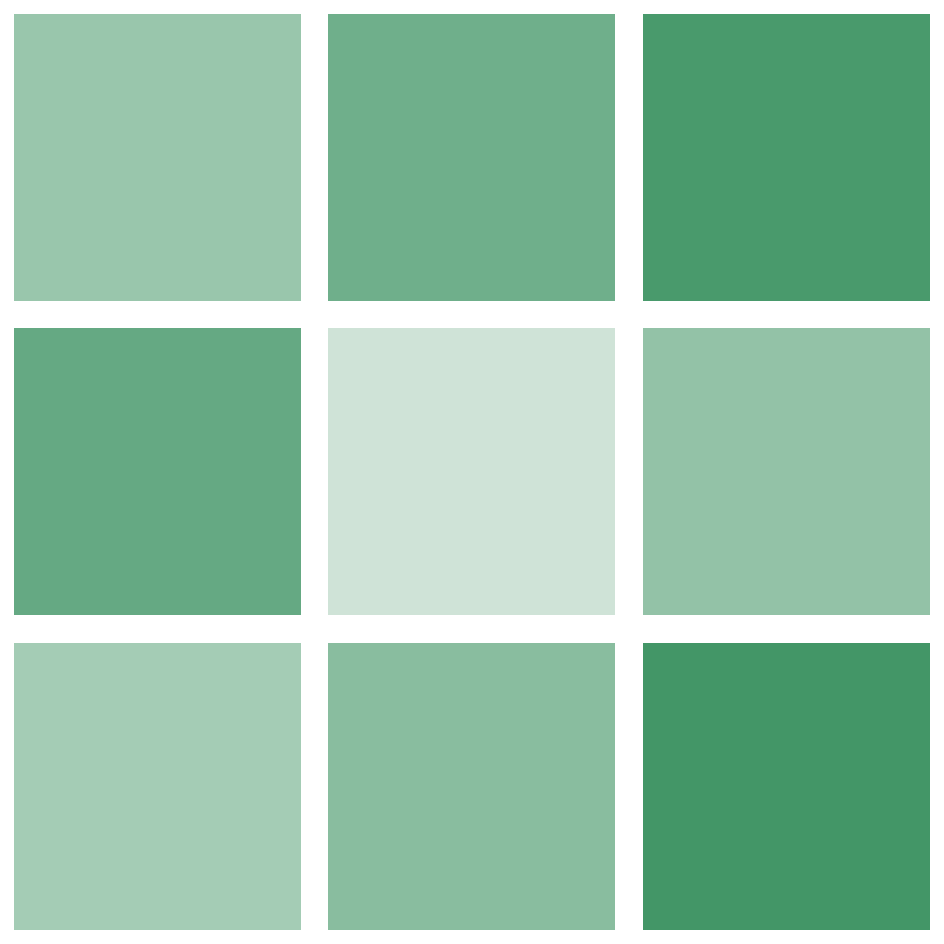}} &\\[10pt]
 \textsc{explosion} &  \texttt{anger} & 128.9** & \raisebox{-.37\height}{\includegraphics[height=25pt]{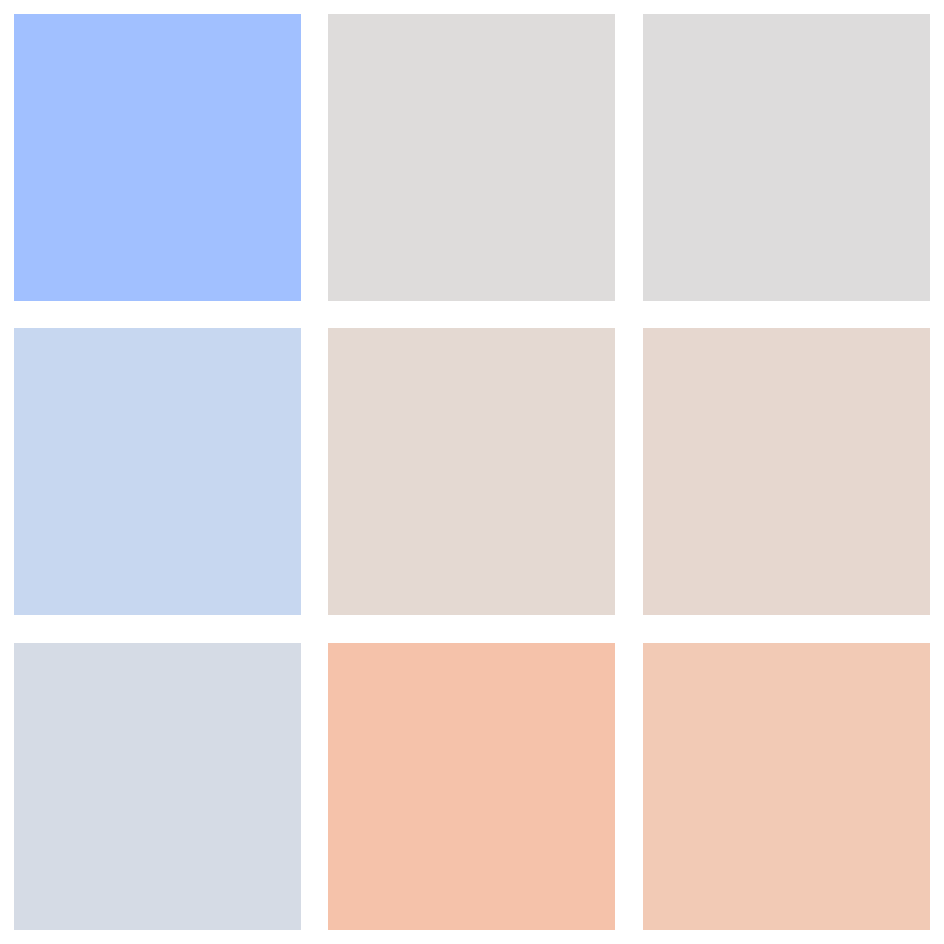}} & & \raisebox{-.37\height}{\includegraphics[height=25pt]{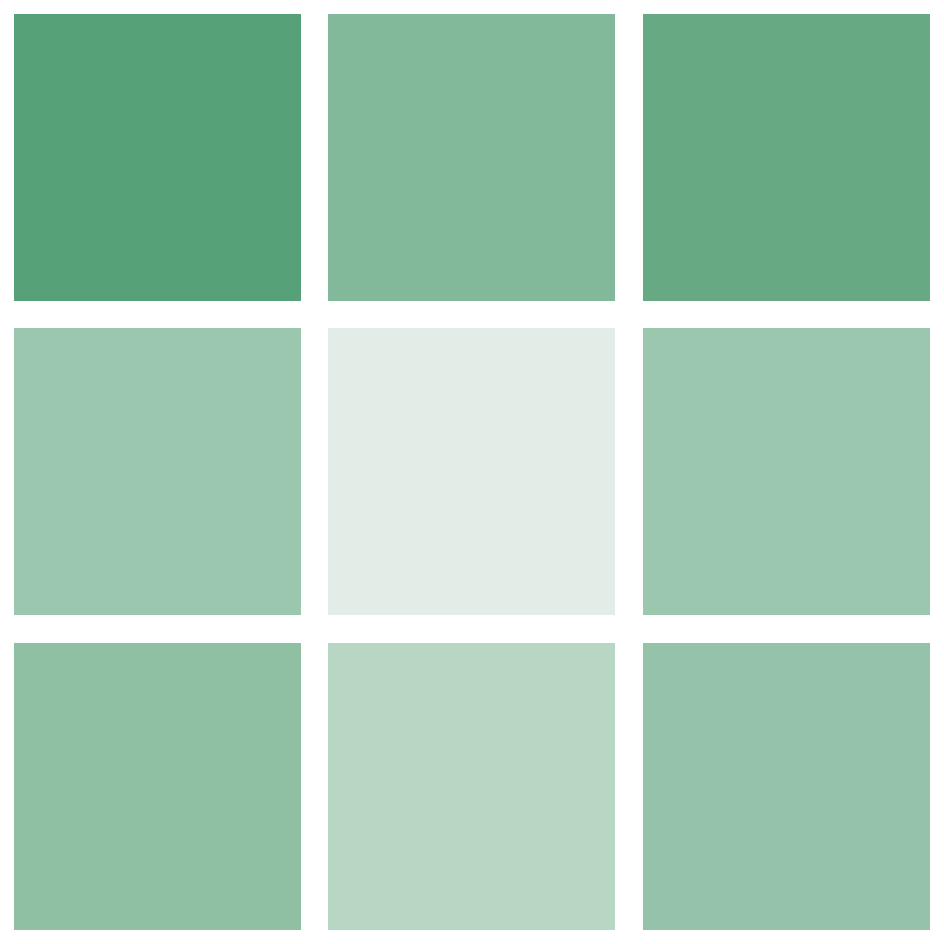}} &\\[10pt]
\textsc{shiver} & \texttt{fear} & 69.0** & \raisebox{-.37\height}{\includegraphics[height=25pt]{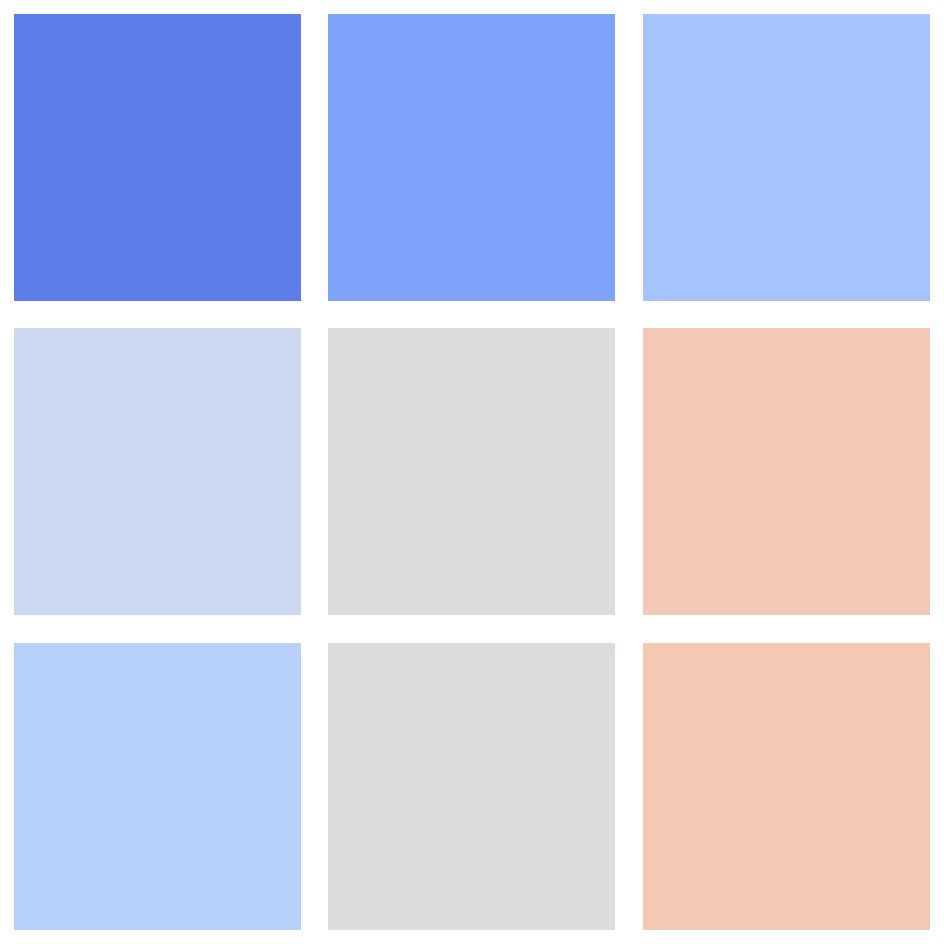}} & & \raisebox{-.37\height}{\includegraphics[height=25pt]{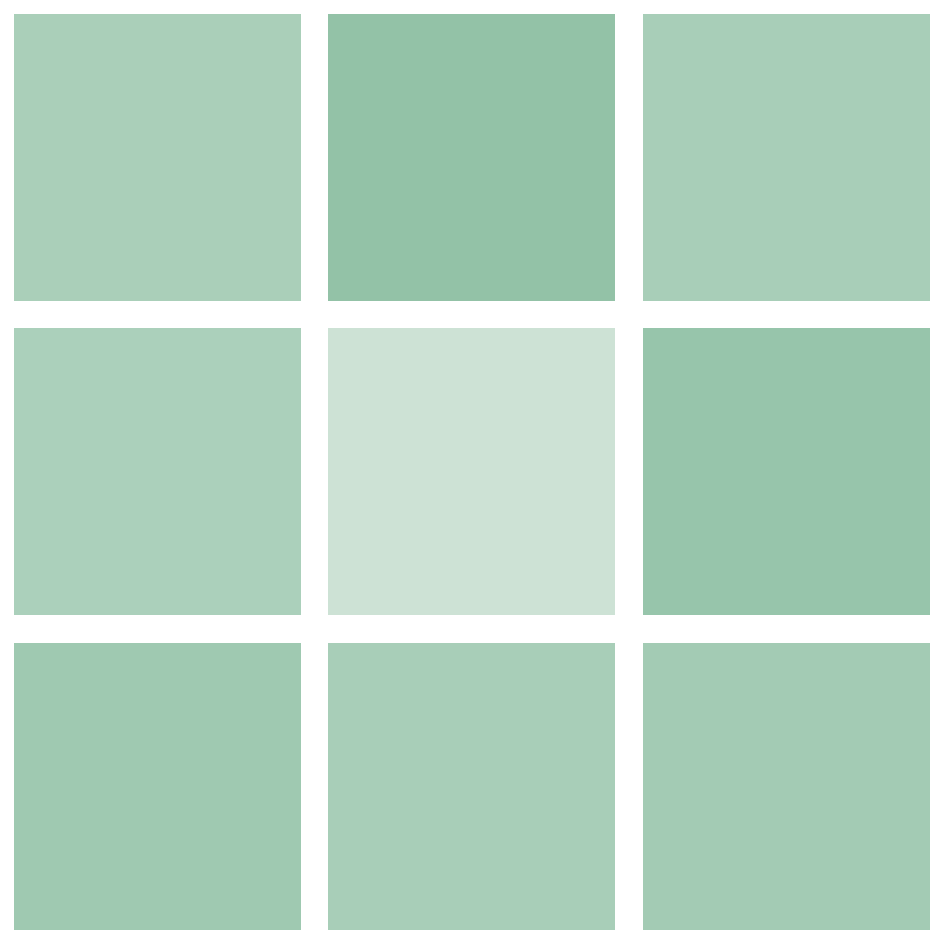}} &\\[10pt]
\bottomrule
\end{tabular}
\label{tab:exp2}
\end{table}
We ran Kruskall-Wallis tests to compare the non-normal distributions of intensity rating for different \abbr{} variants under one animation scheme.
Overall, when testing the 8 variants mutually, we observed significant differences under different speeds and entropy for all four emotions, where each \textit{p} value was less than 0.01.
Furthermore, we examined the pairwise distribution difference, \ie{}, 28 pairs of the speed-entropy parameters for each animation scheme.
The proportion of significantly (significance level $\alpha=0.05$) different pairs were \textsc{dance}--85.7\%, \textsc{fade}--32.4\%, \textsc{explosion}--82.1\%, and \textsc{shiver}--60\%.
Both the \textsc{fade} and \textsc{shiver} schemes adopted the random grouping strategy, and the positional change was smaller than the other two schemes.
These inborn features might contribute to the relatively low distinctions in their parameter pairs, especially when the configurations were close, \eg, low entropy versus medium entropy for medium speed.
This is also reflected in~\autoref{tab:exp1-lorem}, where 15\%$\sim$20\% of people found the animation delivered no emotions.
Notwithstanding, the test results for pairs with the four extremes were significant across four animation schemes.
These results \revv{suggest} that the configuration of entropy and speed could change the perceived emotion intensity of the four design cases.
In addition, few participants expressed disagreement with the presumed emotion in Study II.
Among all 2,188 valid paired comparisons, there were 56 objections, making a proportion of 2.56\%.
This implicit crowd agreement \revv{suggests} that parameter combinations other than medium speed and entropy used in Study I might deliver the intended emotion.

\section{User Study}
\label{sec:user-study}
\draft{We conducted a user study to evaluate (i) how the general public would perceive, assess, and leverage \abbr{}, and (ii) how well the composite approach fit in the intended usage scenario.
The evaluation was based on a prototype authoring tool implementing the four design cases using the composite approach.
We conducted an interview study with an \abbr{} creation task.}

\subsection{Prototype Overview}
\begin{figure*}[ht]
\centering
  \includegraphics[width=\textwidth]{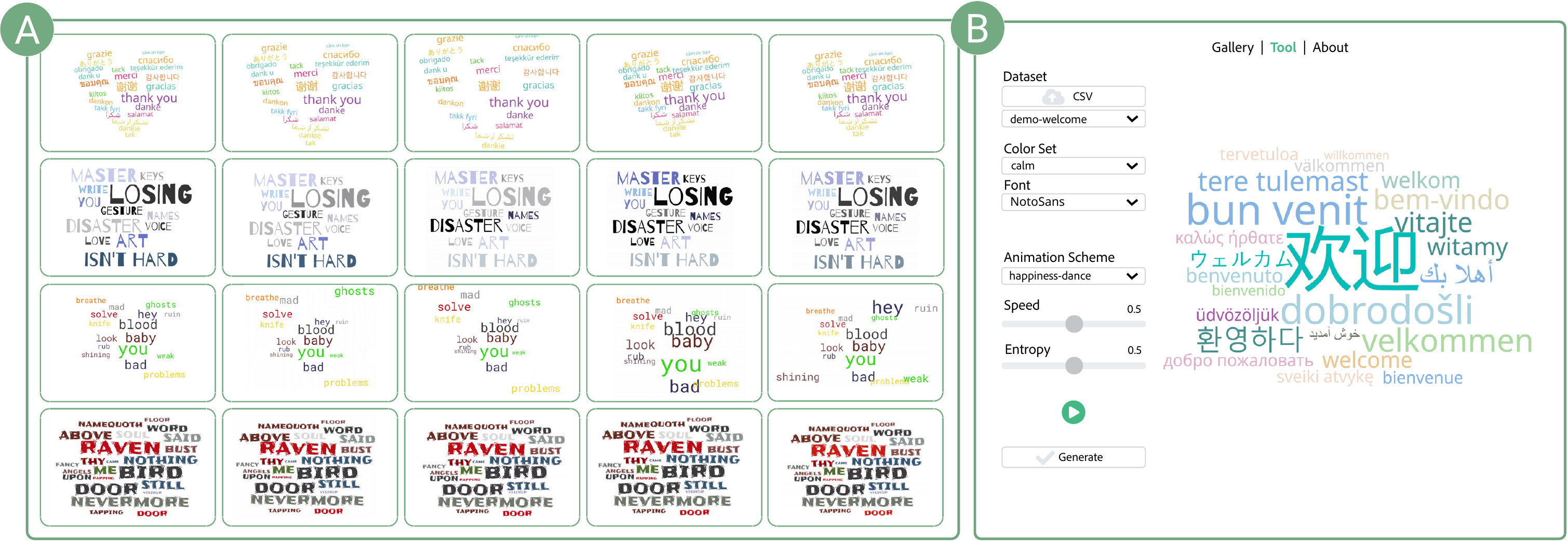}%
  \caption{(A) A gallery of \abbr{} examples with the four animation schemes: \textsc{dance}, \textsc{fade}, \textsc{explosion}, and \textsc{shiver}. (B) The interface of an \abbr{} creation tool.}
  \label{fig:prototype}
  \vspace*{-\baselineskip}
\end{figure*} 
The prototype can be accessed at \demo{}.
\autoref{fig:prototype} shows the user interface.
The input is CSV-based wordle data containing (text, weight) tuples.
There are three steps to customizing an \abbr{}, where each user operation will take effect in the central canvas.
First, apart from the default settings, users can decide on its static appearance by selecting an emotionally congruent color scheme~\cite{Bartram17color} and typeface~\cite{Kucher18affective}.
Second, they select from a template list a dedicated unit animation scheme that matches the basic emotion class that meets their needs.
Then they can fine-tune the animated wordle by sliding the bars for entropy and speed.
Once they are satisfied with the result, they may record it as a video and convert it into a GIF format.

\subsection{Participants}
\label{sec:interview}
We recruited sixteen volunteers (denoted as P1--P16) by disseminating online posters.
Their ages ranged from 20 to 40. There were 11 females and 5 males with diverse backgrounds, including journalism (3), education (2), data science (2), data visualization (2), the media industry (2), design (2), digital art (1), marketing (1), and management (1).
Among them, three had never heard of wordle; eleven understood its encoding; two had created wordles previously for data analysis reports, blog posts, and presentations.

\subsection{Procedure}
The usability study was in remote mode with individual video meetings through four steps.
(1) We first collected the demographic information and introduced the concept of \abbr{}.
(2) We used \abbr{}s in the gallery (\autoref{fig:prototype}) to test whether participants could recognize the conveyed emotion with full visual embellishment.
(3) Then we walked the participants through the system and invited them to generate at least one \abbr{} using their own data or sample data we prepared.
Afterward, participants were required to complete a questionnaire about the system with the listed questions using a 5-point Likert scale (1 for ``strongly disagree'', and 5 for ``strongly agree'').
(4) Followed by the questions, we interviewed participants about their expected usage of \abbr{} and asked for open comments, especially on any potential improvements to the tool and the reasons for their scores. 

\begin{enumerate}
\renewcommand{\labelenumi}{\textbf{Q\theenumi.}}
\renewcommand{\labelenumii}{ Q\theenumi.}


  \item \textbf{Expressiveness} (G1, G3).\textit{``I am satisfied with the final \abbr{} GIF I made, and it delivers my message.''} If the rating is below $5$, participants should demonstrate their anticipation to an ``ideal'' \abbr{}.
  They will clarify which part of the animation is favorable.
    \item \textbf{Convenience} (G2). \textit{``I am satisfied with the simple interaction to tune the wordle.''}
    If participants do not fully agree, we ask them to list the expected functionalities.
    \item \textbf{Intuitiveness} (G1). \textit{``The control of speed and entropy accords with my expectation.''}
    Participants are further asked to illustrate any discord by indicating their original thoughts if they do not score a $5$.
    \item \textbf{Practicality} (G3). \textit{``I would like to use an \abbr{} rather than a static wordle in future data presentation.''} 
    Participants should elaborate on the benefits or the drawbacks of a static wordle over an \abbr{}; We invite them to brainstorm any potential real-life scenarios.

\end{enumerate}

\begin{figure}[t]
\centering
  \includegraphics[width=\columnwidth]{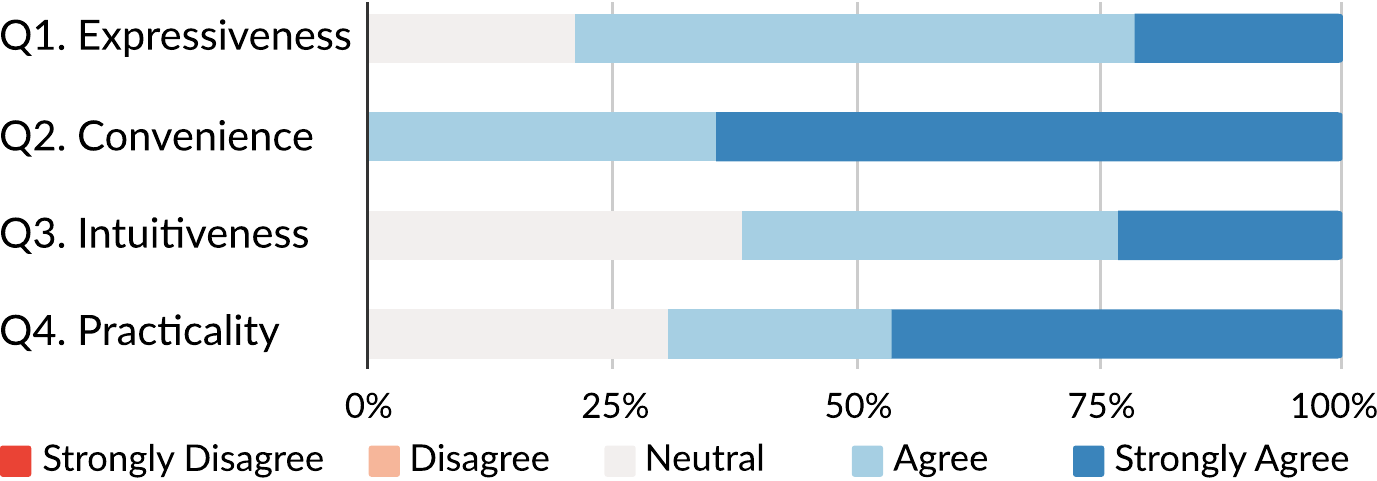}%
  \caption{Distribution of users' feedback on the questionnaire.
  }
  \label{fig:interview}
\end{figure}

\subsection{Feedback}  
\autoref{fig:interview} depicts the quantitative results of the questionnaire, where the prototype system is well recognized for its expressiveness, convenience, intuitiveness, and practicality.
Below we report the interview feedback in detail.

\textbf{Assessing the generated \abbr{}}.
People were generally positive about the output \abbr{}s from the prototype tool, though it only covered a small portion of the entire design space.
For existing \abbr{}s in the gallery, all participants reported the intended emotion of \abbr{}s, which suggested that animation expressions from various design factors could be combined when they were emotionally congruent.
$79\%$ of the participants were satisfied with the \abbr{}s they created using the system, which they believed to have delivered their intended emotion (Q1).
In terms of personalization, we saw evidence of the participatory culture~\cite{viegas09participatory} in authoring \abbr{}s, where participants enthusiastically configured both the animation parameters and other text properties like the layout and color scheme.

There were some complaints about insufficient choices of animation and limited editorial layers in the current tool.
P8, a video blogger, hoped to combine an \abbr{} with background music.
P12, a UX designer, expressed a strong interest in going into the template implementation details and tweaking the animation with an extra timeline design.
Notwithstanding, our long-term goal is to build up an open repository of animation schemes contributed by the creative community, further enriching the visual vocabulary of \abbr{}.
To meet expectations from people with more design skills, the tool should allow higher agency and finer granularity~\cite{mendez2018agency}.
In addition, the support of external editorial layers~\cite{lan21kineticharts} and interoperability should be considered.

\textbf{Evaluating the authoring process}.
Participants raised no issues with the current setting of selecting a template from a preset.
Besides that, everyone was satisfied with the tuning operations with two sliders for entropy and speed (Q2).
P11 commented \textit{``I cannot design animation myself. Without templates, I might give up using animation... The slider helps me arrive at a suitable place.''}

While some participants adapted quickly to the two parameters, 35\% criticized entropy for not being adequately self-explanatory (Q3).
They explained in the follow-up interviews that they could quickly understand the speed parameter but found it hard to interpret the entropy sometimes.
Some claimed they were unfamiliar with the term, and others found it difficult to conceive the results, as words were randomly grouped in the system to achieve different entropy values.
P4 said \textit{``I cannot conceive what will happen until I actually drag the slider and learn from the result.''}
While exposing only a small number of parameters reduced the decisions that inexperienced users should make, this might potentially sacrifice the parameter intuitiveness and hinder the design iteration.
It remains an open challenge to navigate users or make recommendations in a high-dimensional design space, especially when their ideas for the final result are fuzzy and gradually refined.

\textbf{Envisioning application scenarios.}
According to the participants, \abbr{} is beneficial for engagement and entertainment.
However, it was noted that one should be careful when using animation in a serious context.
Of the participants, $69\%$ expressed willingness to use \abbr{}s in future data presentations (Q4).
The most anticipated application scenarios include presentation and user-content generation (blog, video blog, social sharing).
Notably, participants with a marketing or media background (P1, P3, P14) acknowledged the need for emotion expression in wordle, mentioning their demand for producing emotionally consistent content.
While an \abbr{} allows higher engagement or entertainment level, it might harm the information communication when people are paying more attention to the animation rather than the textual content.
P5, a blogger who used a wordle to summarize his post, was concerned that his core message might be overshadowed by fancy animations.
P7, a primary school teacher, raised a similar concern.
She regularly used animated wordle to show real-time polling results in online teaching.
She commented that \textit{``the animation might break the students' concentration''.}
As a compromise, it might be interesting to combine both the static and animated state by enabling a pause.

\section{Discussion}
\label{sec:discussion}
We reflect on the limitations of our study and recommend future research topics in brief.

\subsection{Limitations}

\ \ \ \   \textbf{Trade-offs in Animation.}
The idea of animating wordles has two inherent pitfalls.
Firstly, when designing an \abbr{}, it becomes almost inevitable to leave some white space between text elements to allow movements.
Therefore, the data-ink ratio~\cite{tufte1985visual} of an \abbr{} is likely lower than a static wordle.
Secondly, people cannot perceive too many minute changes in an animation simultaneously~\cite{Tversky02facilitate}.
This constrains the number of words as well as the play speed in an \abbr{}.
Notwithstanding, these design trade-offs are practical in the context of data communication.

\textbf{Prerequisites of the Composite Approach.}
The controlled experiments with four design cases initially validated the composite approach.
Following the approach, more design cases shall be developed, thereby enhancing the authoring tool for end-users in a casual context.
However, it should be noted that the effectiveness of an \abbr{} is attributed to the backbone animation scheme of a single text element.
In this study, we learned from promising examples and co-designed \abbr{}s with an experienced designer; yet it remains challenging to ensure high crowd agreement over the expressed emotion, as seen from the experiment results.
Notably, it is assumed that only one emotion underlies the wordle data, which may not hold for those made from a document collection.
In light of these limitations, we encourage future explorations of more effective approaches for generating \abbr{}s from the vast design space of the animated text.


\textbf{Threats to Validity.}
First, while the composite approach aims to produce wordle animations matching any self-reported emotions, our design cases are limited in the animation diversity and the emotion coverage (4/20 of the base emotions in the GEW model~\cite{scherer13grid}).
As such, the approach has not been comprehensively evaluated in the follow-up experiments and the user study.
Second, because the controlled experiment adopted stimuli with 18 words of regular lengths, it is unclear whether a dense or complex wordle can offset the identified effects.
Besides that, the experiment results may not generalize to crowds with other cultural backgrounds since we required participants to be fluent in English.
Last, in the user study, we invited participants for a one-time generation of an \abbr{}, where most could produce a satisfying one with the tool.
The difference in the dataset and underlying emotion provided initial evidence of the tool's expressiveness.
However, a long-term evaluation was missing to understand whether the tool remains effective in regular use under dynamic scenarios.

\subsection{Implications \& Future Work}

Our design cases demonstrate that controlling the speed and randomness in animation help express subtle emotions. This in turn can be an expressive pathway to enrich the visual vocabulary in visualizations, moving beyond distinct data-driven channels ~\cite{wood2022beyond,sturdee2022painting}.
A wordle often comprises tens of marks with the unused visual channel of position.
In light of this, our qualitative findings and generation approach may also apply to visualizations with similar features, such as graph and unit visualizations~\cite{park2017atom}.
Notably, whether animation schemes without positional changes may apply to other visualizations yet requires more evaluation.

We foresee the following directions for future studies on affective visualization design and authoring support.

\textit{Gain deeper insights into the design factors for emotion conveyance.}
To understand what contributes to emotion expression, we conducted content analysis on a self-curated dataset of animated text.
Our findings are limited to the dataset coverage and our subjective analysis, we envision that an enlarged corpus with a closer examination of the animation techniques \rev{would} result in more fruitful discoveries.
For instance, future work may include particle animations and study staging in the animation.
In addition, understanding the fundamental construct of emotion delivery remains a critical step towards a more executable taxonomy for affective visualizations for follow-up designs or even automatic generation~\cite{di2023doom,chen2023does}.

\textit{Refine the tool for a smoother creating process.}
Due to the difficulty of inferring the nuanced emotion from pure wordle data, our approach relies on human authors to align the animation with the intended emotion by sliding two continuous axes.
For instance, users can use a fast and chaotic anger scene (\eg{}~\textsc{explosion}) to suggest extreme annoyance.
However, this is sometimes demanding for inexperienced people, where guidance or recommendations on the parameter space can be beneficial (\eg,~\cite{Yuki20SeqGallery,li2022gotreescape}).
For users with higher design expertise and expressibility needs, a more flexible authoring experience is expected for tweaking animation details~\cite{li2021What}.
Besides, with sentiment analysis~\cite{wankhade2022survey}, future studies may explore the automatic generation of \abbr{} under scenarios with richer contextual information and representative emotion classes.
\vspace{-0.5cm}
\section{Conclusion}

\label{sec:conclusion}
We explored the opportunity to animate wordles to deliver their intended emotion and enhance communication.
Informed by the current practice of animated text, we contributed a composite approach for constructing emotional animated wordle, namely \abbr{}.
It \revv{reuses} existing animation schemes for a single text and \revv{extends} to wordle with two parameters, \ie{}, entropy and speed.
Based on the approach, we developed four design cases.
For these four cases, two controlled crowd experiments confirmed that people generally agreed on the emotional expression of the given animation schemes, and the two parameters could fine-tune the emotion intensity.
Moreover, we applied the framework to a prototype authoring tool.
Participants with various backgrounds in our interview study confirmed its expressiveness, convenience, intuitiveness, and practicality.

We call for more attention to the affective aspect of data visualization and hope our work can inspire future research in designs for emotion conveyance in a way that directly speaks to the audience.
We envision that animation design may be adapted into an even more automatic pipeline and extend the vocabulary of visualizations.

\vspace{-0.5cm}

\section*{Acknowledgment}
This research was supported by HKRGC General Research Fund 16210722, the Natural Science Foundation of China (NSFC No.62202105), and Shanghai Municipal Science and Technology  (No. 21ZR1403300 \& No. 21YF1402900).
We thank Dr. Xingyu Lan, Mr. Zhibang Jiang, the interviewees, and the anonymous reviewers for their valuable feedback.
\bibliographystyle{abbrv-doi-hyperref}
\bibliography{reference.bib}

\begin{thebibliography}{10}

\bibitem{aftereffects}
{Adobe Inc.}
\newblock {After Effects}.
\newblock \url{https://www.adobe.com/products/aftereffects.html}, 2022.
\newblock Retrieve on 25-Nov-2022.

\bibitem{anderson2021AffectMap}
C.~L. Anderson and A.~C. Robinson.
\newblock Affective congruence in visualization design: Influences on reading
  categorical maps.
\newblock {\em IEEE Trans. Vis. Comput. Graph.}, 28(8):2687--2878, 2021.
  \href{https://doi.org/10.1109/TVCG.2021.3050118}
{doi: {{%
10\hspace{.1pt}\discretionary{.}{%
}{.}\hspace{.4pt}1109\discretionary{/}{%
}{/}TVCG\hspace{.1pt}\discretionary{.}{%
}{.}\hspace{.4pt}2021\hspace{.1pt}\discretionary{.}{%
}{.}\hspace{.4pt}3050118}}}


\bibitem{aoki2022emoballon}
T.~Aoki, R.~Chujo, K.~Matsui, S.~Choi, and A.~Hautasaari.
\newblock {EmoBalloon}--{C}onveying emotional arousal in text chats with speech
  balloons.
\newblock In {\em Proc.\ of the ACM Conference on Human Factors in Computing
  Systems (CHI)}, pp. 527:1--527:16. ACM, 2022.
  \href{https://doi.org/10.1145/3491102.3501920}
{doi: {{%
10\hspace{.1pt}\discretionary{.}{%
}{.}\hspace{.4pt}1145\discretionary{/}{%
}{/}3491102\hspace{.1pt}\discretionary{.}{%
}{.}\hspace{.4pt}3501920}}}


\bibitem{Bartram2009Parameters}
L.~Bartram and A.~Nakatani.
\newblock {Distinctive Parameters of Expressive Motion}.
\newblock In {\em Computational Aesthetics in Graphics, Visualization, and
  Imaging}, pp. 129--136. The Eurographics Association, 2009.
  \href{https://doi.org/10.2312/COMPAESTH/COMPAESTH09/129-136}
{doi: {{%
10\hspace{.1pt}\discretionary{.}{%
}{.}\hspace{.4pt}2312\discretionary{/}{%
}{/}COMPAESTH\discretionary{/}{%
}{/}COMPAESTH09\discretionary{/}{%
}{/}129\discretionary{%
}{-}{-}136}}}


\bibitem{Bartram17color}
L.~Bartram, A.~Patra, and M.~Stone.
\newblock Affective color in visualization.
\newblock In {\em Proc.\ of the ACM Conference on Human Factors in Computing
  Systems (CHI)}, pp. 1364--–1374. ACM, 2017.
  \href{https://doi.org/10.1145/3025453.3026041}
{doi: {{%
10\hspace{.1pt}\discretionary{.}{%
}{.}\hspace{.4pt}1145\discretionary{/}{%
}{/}3025453\hspace{.1pt}\discretionary{.}{%
}{.}\hspace{.4pt}3026041}}}


\bibitem{brath20VisWithText}
R.~Brath.
\newblock {\em Visualizing with Text}, pp. 42--76.
\newblock A K Peters/CRC Press, Boca Raton, 2020.
  \href{https://doi.org/10.1201/9780429290565}
{doi: {{%
10\hspace{.1pt}\discretionary{.}{%
}{.}\hspace{.4pt}1201\discretionary{/}{%
}{/}9780429290565}}}


\bibitem{chang1998negotiation}
B.-W. Chang, J.~D. Mackinlay, P.~T. Zellweger, and T.~Igarashi.
\newblock A negotiation architecture for fluid documents.
\newblock In {\em Proc.\ of the ACM Symposium on User Interface Software and
  Technology (UIST)}, pp. 123--132. ACM, 1998.
  \href{https://doi.org/10.1145/288392.288585}
{doi: {{%
10\hspace{.1pt}\discretionary{.}{%
}{.}\hspace{.4pt}1145\discretionary{/}{%
}{/}288392\hspace{.1pt}\discretionary{.}{%
}{.}\hspace{.4pt}288585}}}


\bibitem{chen2023does}
Q.~Chen, S.~Cao, J.~Wang, and N.~Cao.
\newblock How does automation shape the process of narrative visualization: A
  survey of tools.
\newblock {\em {IEEE} Trans. Vis. Comput. Graph.}, 2023.
\newblock Early Access. \href{https://doi.org/10.1109/TVCG.2023.3261320}
{doi: {{%
10\hspace{.1pt}\discretionary{.}{%
}{.}\hspace{.4pt}1109\discretionary{/}{%
}{/}TVCG\hspace{.1pt}\discretionary{.}{%
}{.}\hspace{.4pt}2023\hspace{.1pt}\discretionary{.}{%
}{.}\hspace{.4pt}3261320}}}


\bibitem{chevalier2010using}
F.~Chevalier, P.~Dragicevic, A.~Bezerianos, and J.-D. Fekete.
\newblock Using text animated transitions to support navigation in document
  histories.
\newblock In {\em Proc.\ of the ACM Conference on Human Factors in Computing
  Systems (CHI)}, pp. 683--692, 2010.
  \href{https://doi.org/10.1145/1753326.1753427}
{doi: {{%
10\hspace{.1pt}\discretionary{.}{%
}{.}\hspace{.4pt}1145\discretionary{/}{%
}{/}1753326\hspace{.1pt}\discretionary{.}{%
}{.}\hspace{.4pt}1753427}}}


\bibitem{Chevalier16animation}
F.~Chevalier, N.~H. Riche, C.~Plaisant, A.~Chalbi, and C.~Hurter.
\newblock Animations 25 years later: New roles and opportunities.
\newblock In {\em Proc.\ of the International Working Conference on Advanced
  Visual Interfaces (AVI)}, pp. 280--287. ACM, 2016.
  \href{https://doi.org/10.1145/2909132.2909255}
{doi: {{%
10\hspace{.1pt}\discretionary{.}{%
}{.}\hspace{.4pt}1145\discretionary{/}{%
}{/}2909132\hspace{.1pt}\discretionary{.}{%
}{.}\hspace{.4pt}2909255}}}


\bibitem{chi15morphable}
M.~Chi, S.~Lin, S.~Chen, C.~Lin, and T.~Lee.
\newblock Morphable word clouds for time-varying text data visualization.
\newblock {\em {IEEE} Trans. Vis. Comput. Graph.}, 21(12):1415--1426, 2015.
  \href{https://doi.org/10.1109/TVCG.2015.2440241}
{doi: {{%
10\hspace{.1pt}\discretionary{.}{%
}{.}\hspace{.4pt}1109\discretionary{/}{%
}{/}TVCG\hspace{.1pt}\discretionary{.}{%
}{.}\hspace{.4pt}2015\hspace{.1pt}\discretionary{.}{%
}{.}\hspace{.4pt}2440241}}}


\bibitem{Dang19WordStream}
T.~Dang, H.~N. Nguyen, and V.~Pham.
\newblock {WordStream}: Interactive visualization for topic evolution.
\newblock In {\em Proc. of the Eurographics/IEEE VGTC Conference on
  Visualization: Short Papers}, pp. 103--107. Eurographics Association, 2019.
  \href{https://doi.org/10.2312/evs.20191178}
{doi: {{%
10\hspace{.1pt}\discretionary{.}{%
}{.}\hspace{.4pt}2312\discretionary{/}{%
}{/}evs\hspace{.1pt}\discretionary{.}{%
}{.}\hspace{.4pt}20191178}}}


\bibitem{felecia15textility}
F.~Davis.
\newblock The textility of emotion: A study relating computational textile
  textural expression to emotion.
\newblock In {\em Proc. of the ACM Conference on Creativity and Cognition
  (C\&C)}, pp. 23--–32. {ACM}, 2015.
  \href{https://doi.org/10.1145/2757226.2757231}
{doi: {{%
10\hspace{.1pt}\discretionary{.}{%
}{.}\hspace{.4pt}1145\discretionary{/}{%
}{/}2757226\hspace{.1pt}\discretionary{.}{%
}{.}\hspace{.4pt}2757231}}}


\bibitem{desai19geppetto}
R.~Desai, F.~Anderson, J.~Matejka, S.~Coros, J.~McCann, G.~Fitzmaurice, and
  T.~Grossman.
\newblock Geppetto: Enabling semantic design of expressive robot behaviors.
\newblock In {\em Proc.\ of the ACM Conference on Human Factors in Computing
  Systems (CHI)}, pp. 369:1--369:15. ACM, 2019.
  \href{https://doi.org/10.1145/3290605.3300599}
{doi: {{%
10\hspace{.1pt}\discretionary{.}{%
}{.}\hspace{.4pt}1145\discretionary{/}{%
}{/}3290605\hspace{.1pt}\discretionary{.}{%
}{.}\hspace{.4pt}3300599}}}


\bibitem{desmet2003measuring}
P.~Desmet.
\newblock Measuring emotion: Development and application of an instrument to
  measure emotional responses to products.
\newblock In {\em Funology}, pp. 111--123. Springer, 2003.

\bibitem{di2023doom}
S.~Di~Bartolomeo, V.~Schetinger, J.~L. Adams, A.~M. McNutt, M.~El-Assady, and
  M.~Miller.
\newblock Doom or deliciousness: Challenges and opportunities for visualization
  in the age of generative models.
\newblock {\em Comput. Graph. Forum}, 2023.
  \href{https://doi.org/10.31219/osf.io/3jrcm}
{doi: {{%
10\hspace{.1pt}\discretionary{.}{%
}{.}\hspace{.4pt}31219\discretionary{/}{%
}{/}osf\hspace{.1pt}\discretionary{.}{%
}{.}\hspace{.4pt}io\discretionary{/}{%
}{/}3jrcm}}}


\bibitem{felix17taking}
C.~Felix, S.~Franconeri, and E.~Bertini.
\newblock Taking word clouds apart: An empirical investigation of the design
  space for keyword summaries.
\newblock {\em {IEEE} Trans. Vis. Comput. Graph.}, 24(1):657--666, 2017.
  \href{https://doi.org/10.1109/TVCG.2017.2746018}
{doi: {{%
10\hspace{.1pt}\discretionary{.}{%
}{.}\hspace{.4pt}1109\discretionary{/}{%
}{/}TVCG\hspace{.1pt}\discretionary{.}{%
}{.}\hspace{.4pt}2017\hspace{.1pt}\discretionary{.}{%
}{.}\hspace{.4pt}2746018}}}


\bibitem{Feng17MotionScape}
C.~Feng, L.~Bartram, and D.~Gromala.
\newblock Beyond data: Abstract motionscapes as affective visualization.
\newblock {\em Leonardo}, 50(2):205--206, 2017.
  \href{https://doi.org/10.1162/LEON_a_01229}
{doi: {{%
10\hspace{.1pt}\discretionary{.}{%
}{.}\hspace{.4pt}1162\discretionary{/}{%
}{/}LEON\_a\_01229}}}


\bibitem{Feng14MotionScape}
C.~Feng, L.~Bartram, and B.~E. Riecke.
\newblock Evaluating affective features of {3D} motionscapes.
\newblock In {\em Proc.\ of the ACM Symposium on Applied Perception (SAP)}, pp.
  23--30. ACM, 2014. \href{https://doi.org/10.1145/2628257.2628264}
{doi: {{%
10\hspace{.1pt}\discretionary{.}{%
}{.}\hspace{.4pt}1145\discretionary{/}{%
}{/}2628257\hspace{.1pt}\discretionary{.}{%
}{.}\hspace{.4pt}2628264}}}


\bibitem{fisher10animation}
D.~Fisher.
\newblock Animation for visualization: Opportunities and drawbacks.
\newblock In {\em Beautiful Visualization - {L}ooking at Data Through the Eyes
  of Experts}, pp. 329--351. O'Reilly, 2010.

\bibitem{Forlizzi03Kinedit}
J.~Forlizzi, J.~Lee, and S.~Hudson.
\newblock The kinedit system: Affective messages using dynamic texts.
\newblock In {\em Proc.\ of the ACM Conference on Human Factors in Computing
  Systems (CHI)}, pp. 377--–384. ACM, 2003.
  \href{https://doi.org/10.1145/642611.642677}
{doi: {{%
10\hspace{.1pt}\discretionary{.}{%
}{.}\hspace{.4pt}1145\discretionary{/}{%
}{/}642611\hspace{.1pt}\discretionary{.}{%
}{.}\hspace{.4pt}642677}}}


\bibitem{hearst2020semantic}
M.~A. Hearst, E.~Pedersen, L.~Patil, E.~Lee, P.~Laskowski, and S.~Franconeri.
\newblock An evaluation of semantically grouped word cloud designs.
\newblock {\em {IEEE} Trans. Vis. Comput. Graph.}, 26(9):2748--2761, 2020.
  \href{https://doi.org/10.1109/TVCG.2019.2904683}
{doi: {{%
10\hspace{.1pt}\discretionary{.}{%
}{.}\hspace{.4pt}1109\discretionary{/}{%
}{/}TVCG\hspace{.1pt}\discretionary{.}{%
}{.}\hspace{.4pt}2019\hspace{.1pt}\discretionary{.}{%
}{.}\hspace{.4pt}2904683}}}


\bibitem{Hicke2022wild}
R.~M. Hicke, M.~Goenka, and E.~Alexander.
\newblock Word clouds in the wild.
\newblock In {\em the 7nd Workshop on Visualization for the Digital Humanities
  (Vis4DH)}, 2022.
\newblock arXiv:2210.08059.

\bibitem{Suguru96Multiagent}
S.~Ishizaki.
\newblock Multiagent model of dynamic design: Visualization as an emergent
  behavior of active design agents.
\newblock In {\em Proc. of the ACM Conference on Human Factors in Computing
  Systems (CHI)}, pp. 347--–354. ACM, 1996.
  \href{https://doi.org/10.1145/238386.238566}
{doi: {{%
10\hspace{.1pt}\discretionary{.}{%
}{.}\hspace{.4pt}1145\discretionary{/}{%
}{/}238386\hspace{.1pt}\discretionary{.}{%
}{.}\hspace{.4pt}238566}}}


\bibitem{Jo15WordlePlus}
J.~Jo, B.~Lee, and J.~Seo.
\newblock {WordlePlus}: Expanding wordle's use through natural interaction and
  animation.
\newblock {\em {IEEE} Comput. Graph. \& Appl.}, 35(6):20--28, 2015.
  \href{https://doi.org/10.1109/MCG.2015.113}
{doi: {{%
10\hspace{.1pt}\discretionary{.}{%
}{.}\hspace{.4pt}1109\discretionary{/}{%
}{/}MCG\hspace{.1pt}\discretionary{.}{%
}{.}\hspace{.4pt}2015\hspace{.1pt}\discretionary{.}{%
}{.}\hspace{.4pt}113}}}


\bibitem{Kalra05TextTone}
A.~Kalra and K.~Karahalios.
\newblock {TextTone}: Expressing emotion through text.
\newblock In {\em Proc. of the IFIP Conference on Human-Computer Interaction
  (INTERACT)}, vol. 3585, pp. 966--969. Springer, 2005.
  \href{https://doi.org/10.1007/11555261_81}
{doi: {{%
10\hspace{.1pt}\discretionary{.}{%
}{.}\hspace{.4pt}1007\discretionary{/}{%
}{/}11555261\_81}}}


\bibitem{kennedy18feeling}
H.~Kennedy and R.~L. Hill.
\newblock The feeling of numbers: Emotions in everyday engagements with data
  and their visualisation.
\newblock {\em Sociology}, 52(4):830--848, 2018.
  \href{https://doi.org/10.1177/0038038516674675}
{doi: {{%
10\hspace{.1pt}\discretionary{.}{%
}{.}\hspace{.4pt}1177\discretionary{/}{%
}{/}0038038516674675}}}


\bibitem{Kim2019DataSelfie}
N.~W. Kim, H.~Im, N.~Henry~Riche, A.~Wang, K.~Gajos, and H.~Pfister.
\newblock {DataSelfie}: Empowering people to design personalized visuals to
  represent their data.
\newblock In {\em Proc.\ of the ACM Conference on Human Factors in Computing
  Systems (CHI)}, pp. 79:1--79:12. ACM, 2019.
  \href{https://doi.org/10.1145/3290605.3300309}
{doi: {{%
10\hspace{.1pt}\discretionary{.}{%
}{.}\hspace{.4pt}1145\discretionary{/}{%
}{/}3290605\hspace{.1pt}\discretionary{.}{%
}{.}\hspace{.4pt}3300309}}}


\bibitem{Koh10ManiWordle}
K.~Koh, B.~Lee, B.~H. Kim, and J.~Seo.
\newblock {ManiWordle}: Providing flexible control over wordle.
\newblock {\em {IEEE} Trans. Vis. Comput. Graph.}, 16(6):1190--1197, 2010.
  \href{https://doi.org/10.1109/TVCG.2010.175}
{doi: {{%
10\hspace{.1pt}\discretionary{.}{%
}{.}\hspace{.4pt}1109\discretionary{/}{%
}{/}TVCG\hspace{.1pt}\discretionary{.}{%
}{.}\hspace{.4pt}2010\hspace{.1pt}\discretionary{.}{%
}{.}\hspace{.4pt}175}}}


\bibitem{Yuki20SeqGallery}
Y.~Koyama, I.~Sato, and M.~Goto.
\newblock Sequential gallery for interactive visual design optimization.
\newblock {\em ACM Trans. Graph.}, 39(4):88:1--88:12, 2020.
  \href{https://doi.org/10.1145/3386569.3392444}
{doi: {{%
10\hspace{.1pt}\discretionary{.}{%
}{.}\hspace{.4pt}1145\discretionary{/}{%
}{/}3386569\hspace{.1pt}\discretionary{.}{%
}{.}\hspace{.4pt}3392444}}}


\bibitem{krcadinac2015textual}
U.~Krcadinac, J.~Jovanovic, V.~Devedzic, and P.~Pasquier.
\newblock Textual affect communication and evocation using abstract generative
  visuals.
\newblock {\em IEEE Trans. Hum. Mach. Syst.}, 46(3):370--379, 2015.
  \href{https://doi.org/10.1109/THMS.2015.2504081}
{doi: {{%
10\hspace{.1pt}\discretionary{.}{%
}{.}\hspace{.4pt}1109\discretionary{/}{%
}{/}THMS\hspace{.1pt}\discretionary{.}{%
}{.}\hspace{.4pt}2015\hspace{.1pt}\discretionary{.}{%
}{.}\hspace{.4pt}2504081}}}


\bibitem{Kucher18affective}
K.~Kucher, C.~Paradis, and A.~Kerren.
\newblock The state of the art in sentiment visualization.
\newblock {\em Comput. Graph. Forum}, 37(1):71--96, 2018.
  \href{https://doi.org/10.1111/cgf.13217}
{doi: {{%
10\hspace{.1pt}\discretionary{.}{%
}{.}\hspace{.4pt}1111\discretionary{/}{%
}{/}cgf\hspace{.1pt}\discretionary{.}{%
}{.}\hspace{.4pt}13217}}}


\bibitem{Kulahcioglu20affect}
T.~Kulahcioglu and G.~de~Melo.
\newblock Affect-aware word clouds.
\newblock {\em {ACM} Trans. Interact. Intell. Syst.}, 10(4):34:1--34:25, 2020.
  \href{https://doi.org/10.1145/3370928}
{doi: {{%
10\hspace{.1pt}\discretionary{.}{%
}{.}\hspace{.4pt}1145\discretionary{/}{%
}{/}3370928}}}


\bibitem{lan21kineticharts}
X.~Lan, Y.~Shi, Y.~Wu, X.~Jiao, and N.~Cao.
\newblock Kineticharts: Augmenting affective expressiveness of charts in data
  stories with animation design.
\newblock {\em {IEEE} Trans. Vis. Comput. Graph.}, 28:933--943, 2021.
  \href{https://doi.org/10.1109/TVCG.2021.3114775}
{doi: {{%
10\hspace{.1pt}\discretionary{.}{%
}{.}\hspace{.4pt}1109\discretionary{/}{%
}{/}TVCG\hspace{.1pt}\discretionary{.}{%
}{.}\hspace{.4pt}2021\hspace{.1pt}\discretionary{.}{%
}{.}\hspace{.4pt}3114775}}}


\bibitem{Lan21Smile}
X.~Lan, Y.~Shi, Y.~Zhang, and N.~Cao.
\newblock Smile or scowl? {L}ooking at infographic design through the affective
  lens.
\newblock {\em IEEE Trans. Vis. Comput. Graph.}, 27(2):1095--1105, 2021.
  \href{https://doi.org/10.1109/TVCG.2020.3030435}
{doi: {{%
10\hspace{.1pt}\discretionary{.}{%
}{.}\hspace{.4pt}1109\discretionary{/}{%
}{/}TVCG\hspace{.1pt}\discretionary{.}{%
}{.}\hspace{.4pt}2020\hspace{.1pt}\discretionary{.}{%
}{.}\hspace{.4pt}3030435}}}


\bibitem{lan2022chart}
X.~Lan, Y.~Wu, Q.~Chen, and N.~Cao.
\newblock The chart excites me! {E}xploring how data visualization design
  influences affective arousal.
\newblock {\em arXiv:2211.03296}, 2022.

\bibitem{Lan2022Negative}
X.~Lan, Y.~Wu, Y.~Shi, Q.~Chen, and N.~Cao.
\newblock Negative emotions, positive outcomes? {E}xploring the communication
  of negativity in serious data stories.
\newblock In {\em Proc.\ of the ACM Conference on Human Factors in Computing
  Systems (CHI)}, pp. 28:1--28:14. ACM, 2022.
  \href{https://doi.org/10.1145/3491102.3517530}
{doi: {{%
10\hspace{.1pt}\discretionary{.}{%
}{.}\hspace{.4pt}1145\discretionary{/}{%
}{/}3491102\hspace{.1pt}\discretionary{.}{%
}{.}\hspace{.4pt}3517530}}}


\bibitem{Lee10SparkClouds}
B.~Lee, N.~H. Riche, A.~K. Karlson, and S.~Carpendale.
\newblock {SparkClouds}: Visualizing trends in tag clouds.
\newblock {\em {IEEE} Trans. Vis. Comput. Graph.}, 16(6):1182--1189, 2010.
  \href{https://doi.org/10.1109/TVCG.2010.194}
{doi: {{%
10\hspace{.1pt}\discretionary{.}{%
}{.}\hspace{.4pt}1109\discretionary{/}{%
}{/}TVCG\hspace{.1pt}\discretionary{.}{%
}{.}\hspace{.4pt}2010\hspace{.1pt}\discretionary{.}{%
}{.}\hspace{.4pt}194}}}


\bibitem{Lee07EmotiveCaptioning}
D.~G. Lee, D.~I. Fels, and J.~P. Udo.
\newblock Emotive captioning.
\newblock {\em ACM Comput. Entertain.}, 5(2):11:1--11:15, 2007.
  \href{https://doi.org/10.1145/1279540.1279551}
{doi: {{%
10\hspace{.1pt}\discretionary{.}{%
}{.}\hspace{.4pt}1145\discretionary{/}{%
}{/}1279540\hspace{.1pt}\discretionary{.}{%
}{.}\hspace{.4pt}1279551}}}


\bibitem{Lee06Kinetic}
J.~Lee, S.~Jun, J.~Forlizzi, and S.~E. Hudson.
\newblock Using kinetic typography to convey emotion in text-based
  interpersonal communication.
\newblock In {\em Proc. of the ACM Conference on Designing Interactive Systems
  (DIS)}, pp. 41--49. {ACM}, 2006.
  \href{https://doi.org/10.1145/1142405.1142414}
{doi: {{%
10\hspace{.1pt}\discretionary{.}{%
}{.}\hspace{.4pt}1145\discretionary{/}{%
}{/}1142405\hspace{.1pt}\discretionary{.}{%
}{.}\hspace{.4pt}1142414}}}


\bibitem{Lee02KineticTypographyEngine}
J.~C. Lee, J.~Forlizzi, and S.~E. Hudson.
\newblock The kinetic typography engine: an extensible system for animating
  expressive text.
\newblock In {\em Proc. of the ACM Symposium on User Interface Software and
  Technology (UIST)}, pp. 81--–90. {ACM}, 2002.
  \href{https://doi.org/10.1145/571985.571997}
{doi: {{%
10\hspace{.1pt}\discretionary{.}{%
}{.}\hspace{.4pt}1145\discretionary{/}{%
}{/}571985\hspace{.1pt}\discretionary{.}{%
}{.}\hspace{.4pt}571997}}}


\bibitem{lee2022affective}
E.~Lee-Robbins and E.~Adar.
\newblock Affective learning objectives for communicative visualizations.
\newblock {\em {IEEE} Trans. Vis. Comput. Graph.}, 29:1--11, 2023.
  \href{https://doi.org/10.1109/TVCG.2022.3209500}
{doi: {{%
10\hspace{.1pt}\discretionary{.}{%
}{.}\hspace{.4pt}1109\discretionary{/}{%
}{/}TVCG\hspace{.1pt}\discretionary{.}{%
}{.}\hspace{.4pt}2022\hspace{.1pt}\discretionary{.}{%
}{.}\hspace{.4pt}3209500}}}


\bibitem{li2022gotreescape}
G.~Li and X.~Yuan.
\newblock {GoTreeScape}: Navigate and explore the tree visualization design
  space.
\newblock {\em {IEEE} Trans. Vis. Comput. Graph.}, 2022.
\newblock Early Access. \href{https://doi.org/10.1109/TVCG.2022.3215070}
{doi: {{%
10\hspace{.1pt}\discretionary{.}{%
}{.}\hspace{.4pt}1109\discretionary{/}{%
}{/}TVCG\hspace{.1pt}\discretionary{.}{%
}{.}\hspace{.4pt}2022\hspace{.1pt}\discretionary{.}{%
}{.}\hspace{.4pt}3215070}}}


\bibitem{li2021What}
J.~Li, S.~Hashim, and J.~Jacobs.
\newblock What we can learn from visual artists about software development.
\newblock In {\em Proc.\ of the ACM Conference on Human Factors in Computing
  Systems (CHI)}, pp. 314:1--314:14. ACM, 2021.
  \href{https://doi.org/10.1145/3411764.3445682}
{doi: {{%
10\hspace{.1pt}\discretionary{.}{%
}{.}\hspace{.4pt}1145\discretionary{/}{%
}{/}3411764\hspace{.1pt}\discretionary{.}{%
}{.}\hspace{.4pt}3445682}}}


\bibitem{li2023geocamera}
W.~Li, Z.~Wang, Y.~Wang, D.~Weng, L.~Xie, S.~Chen, H.~Zhang, and H.~Qu.
\newblock Geocamera: Telling stories in geographic visualizations with camera
  movements.
\newblock In {\em Proc.\ of the ACM Conference on Human Factors in Computing
  Systems (CHI)}, pp. 170:1--170:15. ACM, 2023.
  \href{https://doi.org/10.1145/3544548.3581470}
{doi: {{%
10\hspace{.1pt}\discretionary{.}{%
}{.}\hspace{.4pt}1145\discretionary{/}{%
}{/}3544548\hspace{.1pt}\discretionary{.}{%
}{.}\hspace{.4pt}3581470}}}


\bibitem{lorem}
{lipsum.com}.
\newblock Lorem ipsum.
\newblock \url{https://www.lipsum.com/}, 2022.
\newblock Retrieve on 25-Nov-2022.

\bibitem{ma2021review}
H.~Ma and S.~Yarosh.
\newblock A review of affective computing research based on
  function--component--representation framework.
\newblock {\em IEEE Trans. Affect. Comput.}, 2021.
\newblock Early Access. \href{https://doi.org/10.1109/TAFFC.2021.3104512}
{doi: {{%
10\hspace{.1pt}\discretionary{.}{%
}{.}\hspace{.4pt}1109\discretionary{/}{%
}{/}TAFFC\hspace{.1pt}\discretionary{.}{%
}{.}\hspace{.4pt}2021\hspace{.1pt}\discretionary{.}{%
}{.}\hspace{.4pt}3104512}}}


\bibitem{Maharik11Micrography}
R.~Maharik, M.~Bessmeltsev, A.~Sheffer, A.~Shamir, and N.~Carr.
\newblock Digital micrography.
\newblock {\em ACM Trans. Graph.}, 30(4):100:1--100:12, 2011.
  \href{https://doi.org/10.1145/1964921.1964995}
{doi: {{%
10\hspace{.1pt}\discretionary{.}{%
}{.}\hspace{.4pt}1145\discretionary{/}{%
}{/}1964921\hspace{.1pt}\discretionary{.}{%
}{.}\hspace{.4pt}1964995}}}


\bibitem{malik09communicating}
S.~Malik, J.~Aitken, and J.~K. Waalen.
\newblock Communicating emotion with animated text.
\newblock {\em Visual Commun.}, 8(4):469--479, 2009.
  \href{https://doi.org/10.1177/1470357209343375}
{doi: {{%
10\hspace{.1pt}\discretionary{.}{%
}{.}\hspace{.4pt}1177\discretionary{/}{%
}{/}1470357209343375}}}


\bibitem{mendez2018agency}
G.~G. M\'{e}ndez, M.~A. Nacenta, and U.~Hinrichs.
\newblock Considering agency and data granularity in the design of
  visualization tools.
\newblock In {\em Proc.\ of the ACM Conference on Human Factors in Computing
  Systems (CHI)}, pp. 638:1--638:14. ACM, 2018.
  \href{https://doi.org/10.1145/3173574.3174212}
{doi: {{%
10\hspace{.1pt}\discretionary{.}{%
}{.}\hspace{.4pt}1145\discretionary{/}{%
}{/}3173574\hspace{.1pt}\discretionary{.}{%
}{.}\hspace{.4pt}3174212}}}


\bibitem{Minakuchi05Kinetic}
M.~Minakuchi and K.~Tanaka.
\newblock Automatic kinetic typography composer.
\newblock In {\em Proc. of the ACM Conference on Advances in Computer
  Entertainment Technology (ACE)}, pp. 221--224. ACM, 2005.
  \href{https://doi.org/10.1145/1178477.1178512}
{doi: {{%
10\hspace{.1pt}\discretionary{.}{%
}{.}\hspace{.4pt}1145\discretionary{/}{%
}{/}1178477\hspace{.1pt}\discretionary{.}{%
}{.}\hspace{.4pt}1178512}}}


\bibitem{odonovan14font}
P.~O'Donovan, J.~Libeks, A.~Agarwala, and A.~Hertzmann.
\newblock Exploratory font selection using crowdsourced attributes.
\newblock {\em {ACM} Trans. Graph.}, 33(4):92:1--92:9, 2014.
  \href{https://doi.org/10.1145/2601097.2601110}
{doi: {{%
10\hspace{.1pt}\discretionary{.}{%
}{.}\hspace{.4pt}1145\discretionary{/}{%
}{/}2601097\hspace{.1pt}\discretionary{.}{%
}{.}\hspace{.4pt}2601110}}}


\bibitem{ortony1990cognitive}
A.~Ortony, G.~L. Clore, and A.~Collins.
\newblock {\em {The Cognitive Structure of Emotions}}.
\newblock Cambridge University Press, 1990.
  \href{https://doi.org/10.1017/CBO9780511571299}
{doi: {{%
10\hspace{.1pt}\discretionary{.}{%
}{.}\hspace{.4pt}1017\discretionary{/}{%
}{/}CBO9780511571299}}}


\bibitem{park2017atom}
D.~Park, S.~M. Drucker, R.~Fernandez, and N.~Elmqvist.
\newblock Atom: A grammar for unit visualizations.
\newblock {\em {IEEE} Trans. Vis. Comput. Graph.}, 24(12):3032--3043, 2017.
  \href{https://doi.org/10.1109/TVCG.2017.2785807}
{doi: {{%
10\hspace{.1pt}\discretionary{.}{%
}{.}\hspace{.4pt}1109\discretionary{/}{%
}{/}TVCG\hspace{.1pt}\discretionary{.}{%
}{.}\hspace{.4pt}2017\hspace{.1pt}\discretionary{.}{%
}{.}\hspace{.4pt}2785807}}}


\bibitem{luxos}
{Pixar Animation Studios}.
\newblock \url{https://youtu.be/X3AcXraOW_k}, 2022.
\newblock 1:27--1:31, Retrieve on 25-Nov-2022.

\bibitem{pousman2007casual}
Z.~Pousman, J.~T. Stasko, and M.~Mateas.
\newblock Casual information visualization: Depictions of data in everyday
  life.
\newblock {\em {IEEE} Trans. Vis. Comput. Graph.}, 13(6):1145--1152, 2007.
  \href{https://doi.org/10.1109/TVCG.2007.70541}
{doi: {{%
10\hspace{.1pt}\discretionary{.}{%
}{.}\hspace{.4pt}1109\discretionary{/}{%
}{/}TVCG\hspace{.1pt}\discretionary{.}{%
}{.}\hspace{.4pt}2007\hspace{.1pt}\discretionary{.}{%
}{.}\hspace{.4pt}70541}}}


\bibitem{tag2007rivadeneira}
A.~W. Rivadeneira, D.~M. Gruen, M.~J. Muller, and D.~R. Millen.
\newblock Getting our head in the clouds: Toward evaluation studies of
  tagclouds.
\newblock In {\em Proc.\ of the ACM Conference on Human Factors in Computing
  Systems (CHI)}, pp. 995--998. ACM, 2007.
  \href{https://doi.org/10.1145/1240624.1240775}
{doi: {{%
10\hspace{.1pt}\discretionary{.}{%
}{.}\hspace{.4pt}1145\discretionary{/}{%
}{/}1240624\hspace{.1pt}\discretionary{.}{%
}{.}\hspace{.4pt}1240775}}}


\bibitem{russell80circumplex}
J.~A. Russell.
\newblock A circumplex model of affect.
\newblock {\em Journal of Personality and Social Psychology}, 39(6):1161--1178,
  1980. \href{https://doi.org/10.1017/S0954579405050340}
{doi: {{%
10\hspace{.1pt}\discretionary{.}{%
}{.}\hspace{.4pt}1017\discretionary{/}{%
}{/}S0954579405050340}}}


\bibitem{scherer13grid}
K.~R. Scherer, V.~Shuman, J.~Fontaine, and C.~Soriano~Salinas.
\newblock The {GRID} meets the wheel: Assessing emotional feeling via
  self-report.
\newblock {\em Components of emotional meaning: A sourcebook}, pp. 281--298,
  2013. \href{https://doi.org/10.1093/acprof:oso/9780199592746.003.0019}
{doi: {{%
10\hspace{.1pt}\discretionary{.}{%
}{.}\hspace{.4pt}1093\discretionary{/}{%
}{/}acprof\discretionary{:}{%
}{:}oso\discretionary{/}{%
}{/}9780199592746\hspace{.1pt}\discretionary{.}{%
}{.}\hspace{.4pt}003\hspace{.1pt}\discretionary{.}{%
}{.}\hspace{.4pt}0019}}}


\bibitem{roslingifier2022shin}
M.~Shin, J.~Kim, Y.~Han, L.~Xie, M.~Whitelaw, B.~C. Kwon, S.~Ko, and
  N.~Elmqvist.
\newblock Roslingifier: Semi-automated storytelling for animated scatterplots.
\newblock {\em {IEEE} Trans. Vis. Comput. Graph.}, 29(6):2980--2995, 2023.
  \href{https://doi.org/10.1109/TVCG.2022.3146329}
{doi: {{%
10\hspace{.1pt}\discretionary{.}{%
}{.}\hspace{.4pt}1109\discretionary{/}{%
}{/}TVCG\hspace{.1pt}\discretionary{.}{%
}{.}\hspace{.4pt}2022\hspace{.1pt}\discretionary{.}{%
}{.}\hspace{.4pt}3146329}}}


\bibitem{Shu21Gif}
X.~Shu, A.~Wu, J.~Tang, B.~Bach, Y.~Wu, and H.~Qu.
\newblock What makes a {Data-GIF} understandable?
\newblock {\em IEEE Trans. Vis. Comput. Graph.}, 27(02):1492--1502, 2021.
  \href{https://doi.org/10.1109/TVCG.2020.3030396}
{doi: {{%
10\hspace{.1pt}\discretionary{.}{%
}{.}\hspace{.4pt}1109\discretionary{/}{%
}{/}TVCG\hspace{.1pt}\discretionary{.}{%
}{.}\hspace{.4pt}2020\hspace{.1pt}\discretionary{.}{%
}{.}\hspace{.4pt}3030396}}}


\bibitem{shu20dancingword}
X.~Shu, J.~Wu, X.~Wu, H.~Liang, W.~Cui, Y.~Wu, and H.~Qu.
\newblock {DancingWords}: Exploring animated word clouds to tell stories.
\newblock {\em J. Vis.}, 24(1):85--100, 2021.
  \href{https://doi.org/10.1007/s12650-020-00689-0}
{doi: {{%
10\hspace{.1pt}\discretionary{.}{%
}{.}\hspace{.4pt}1007\discretionary{/}{%
}{/}s12650\discretionary{%
}{-}{-}020\discretionary{%
}{-}{-}00689\discretionary{%
}{-}{-}0}}}


\bibitem{stone2004type}
R.~B. Stone, D.~P. Alenquer, and J.~Borisch.
\newblock Type, motion and emotion: A visual amplification of meaning.
\newblock In {\em Design and Emotion}, pp. 212--219. Taylor {\&} Francis,
  London, 2004. \href{https://doi.org/10.1201/9780203608173}
{doi: {{%
10\hspace{.1pt}\discretionary{.}{%
}{.}\hspace{.4pt}1201\discretionary{/}{%
}{/}9780203608173}}}


\bibitem{sturdee2022painting}
M.~Sturdee, S.~Knudsen, and S.~Carpendale.
\newblock Data-painting: Expressive free-form visualisation.
\newblock In {\em Proc.\ of Design Research Society (DRS)}, pp. 38:1--38:16,
  2022. \href{https://doi.org/10.21606/drs.2022.257}
{doi: {{%
10\hspace{.1pt}\discretionary{.}{%
}{.}\hspace{.4pt}21606\discretionary{/}{%
}{/}drs\hspace{.1pt}\discretionary{.}{%
}{.}\hspace{.4pt}2022\hspace{.1pt}\discretionary{.}{%
}{.}\hspace{.4pt}257}}}


\bibitem{thomas1995illusion}
F.~Thomas, O.~Johnston, and F.~Thomas.
\newblock {\em {The illusion of life: Disney animation}}.
\newblock Hyperion New York, 1995.

\bibitem{tufte1985visual}
E.~R. Tufte.
\newblock {\em {The Visual Display of Quantitative Information}}.
\newblock Graphics Press, 1983.

\bibitem{Tversky02facilitate}
B.~Tversky, J.~B. Morrison, and M.~B{\'{e}}trancourt.
\newblock Animation: Can it facilitate?
\newblock {\em Int. J. Hum. Comput. Stud.}, 57(4):247--262, 2002.
  \href{https://doi.org/10.1006/ijhc.2002.1017}
{doi: {{%
10\hspace{.1pt}\discretionary{.}{%
}{.}\hspace{.4pt}1006\discretionary{/}{%
}{/}ijhc\hspace{.1pt}\discretionary{.}{%
}{.}\hspace{.4pt}2002\hspace{.1pt}\discretionary{.}{%
}{.}\hspace{.4pt}1017}}}


\bibitem{urquhart17emotive}
L.~W.~R. Urquhart, A.~Wodehouse, et~al.
\newblock The emotive qualities of patterns: Insights for design.
\newblock In {\em Proc.\ of the 21st International Conference on Engineering
  Design (ICED) Vol 8: Human Behaviour in Design}, pp. 109--118, 2017.

\bibitem{van12design}
T.~Van~Gorp and E.~Adams.
\newblock {\em {Design for Emotion}}.
\newblock Elsevier, 2012.

\bibitem{viegas09participatory}
F.~B. Vi{\'{e}}gas, M.~Wattenberg, and J.~Feinberg.
\newblock Participatory visualization with wordle.
\newblock {\em {IEEE} Trans. Vis. Comput. Graph.}, 15(6):1137--1144, 2009.
  \href{https://doi.org/10.1109/TVCG.2009.171}
{doi: {{%
10\hspace{.1pt}\discretionary{.}{%
}{.}\hspace{.4pt}1109\discretionary{/}{%
}{/}TVCG\hspace{.1pt}\discretionary{.}{%
}{.}\hspace{.4pt}2009\hspace{.1pt}\discretionary{.}{%
}{.}\hspace{.4pt}171}}}


\bibitem{Wang04communicate}
H.~Wang, H.~Prendinger, and T.~Igarashi.
\newblock Communicating emotions in online chat using physiological sensors and
  animated text.
\newblock In {\em Extended Abstracts on Human Factors in Computing Systems
  (CHIEA)}, pp. 1171--1174. ACM, 2004.
  \href{https://doi.org/10.1145/985921.986016}
{doi: {{%
10\hspace{.1pt}\discretionary{.}{%
}{.}\hspace{.4pt}1145\discretionary{/}{%
}{/}985921\hspace{.1pt}\discretionary{.}{%
}{.}\hspace{.4pt}986016}}}


\bibitem{Wang18EdiWordle}
Y.~Wang, X.~Chu, C.~Bao, L.~Zhu, O.~Deussen, B.~Chen, and M.~Sedlmair.
\newblock {EdWordle}: Consistency-preserving word cloud editing.
\newblock {\em {IEEE} Trans. Vis. Comput. Graph.}, 24(1):647--656, 2018.
  \href{https://doi.org/10.1109/TVCG.2017.2745859}
{doi: {{%
10\hspace{.1pt}\discretionary{.}{%
}{.}\hspace{.4pt}1109\discretionary{/}{%
}{/}TVCG\hspace{.1pt}\discretionary{.}{%
}{.}\hspace{.4pt}2017\hspace{.1pt}\discretionary{.}{%
}{.}\hspace{.4pt}2745859}}}


\bibitem{wang20shapewordle}
Y.~Wang, X.~Chu, K.~Zhang, C.~Bao, X.~Li, J.~Zhang, C.~Fu, C.~Hurter, B.~Lee,
  and O.~Deussen.
\newblock {ShapeWordle}: Tailoring wordles using shape-aware archimedean
  spirals.
\newblock {\em {IEEE} Trans. Vis. Comput. Graph.}, 26(1):991--1000, 2020.
  \href{https://doi.org/10.1109/TVCG.2019.2934783}
{doi: {{%
10\hspace{.1pt}\discretionary{.}{%
}{.}\hspace{.4pt}1109\discretionary{/}{%
}{/}TVCG\hspace{.1pt}\discretionary{.}{%
}{.}\hspace{.4pt}2019\hspace{.1pt}\discretionary{.}{%
}{.}\hspace{.4pt}2934783}}}


\bibitem{wang19value}
Y.~Wang, A.~Segal, R.~L. Klatzky, D.~F. Keefe, P.~Isenberg, J.~Hurtienne,
  E.~Hornecker, T.~Dwyer, S.~Barrass, and T.~Rhyne.
\newblock An emotional response to the value of visualization.
\newblock {\em {IEEE} Comput. Graph. and Appl.}, 39(5):8--17, 2019.
  \href{https://doi.org/10.1109/MCG.2019.2923483}
{doi: {{%
10\hspace{.1pt}\discretionary{.}{%
}{.}\hspace{.4pt}1109\discretionary{/}{%
}{/}MCG\hspace{.1pt}\discretionary{.}{%
}{.}\hspace{.4pt}2019\hspace{.1pt}\discretionary{.}{%
}{.}\hspace{.4pt}2923483}}}


\bibitem{wankhade2022survey}
M.~Wankhade, A.~C.~S. Rao, and C.~Kulkarni.
\newblock A survey on sentiment analysis methods, applications, and challenges.
\newblock {\em Artificial Intelligence Review}, pp. 5731--5780, 2022.
  \href{https://doi.org/10.1007/s10462-022-10144-1}
{doi: {{%
10\hspace{.1pt}\discretionary{.}{%
}{.}\hspace{.4pt}1007\discretionary{/}{%
}{/}s10462\discretionary{%
}{-}{-}022\discretionary{%
}{-}{-}10144\discretionary{%
}{-}{-}1}}}


\bibitem{wood2022beyond}
J.~Wood.
\newblock Beyond the walled garden: A visual essay in five chapters.
\newblock In {\em the alt.vis Workshop (alt.vis)}, 2022.

\bibitem{wordArt}
{WordArts.}
\newblock \url{https://wordart.com/}, 2022.
\newblock Retrieved on 25-Nov-2022.

\bibitem{Wu21LQ2}
A.~Wu, L.~Xie, B.~Lee, Y.~Wang, W.~Cui, and H.~Qu.
\newblock Learning to automate chart layout configurations using crowdsourced
  paired comparison.
\newblock In {\em Proc.\ of the ACM Conference on Human Factors in Computing
  Systems (CHI)}, pp. 14:1--14:13. ACM, 2021.
  \href{https://doi.org/10.1145/3411764.3445179}
{doi: {{%
10\hspace{.1pt}\discretionary{.}{%
}{.}\hspace{.4pt}1145\discretionary{/}{%
}{/}3411764\hspace{.1pt}\discretionary{.}{%
}{.}\hspace{.4pt}3445179}}}


\bibitem{wu11semantic}
Y.~Wu, T.~Provan, F.~Wei, S.~Liu, and K.~Ma.
\newblock Semantic{-}preserving word clouds by seam carving.
\newblock {\em Comput. Graph. Forum}, 30(3):741--750, 2011.
  \href{https://doi.org/10.1111/j.1467-8659.2011.01923.x}
{doi: {{%
10\hspace{.1pt}\discretionary{.}{%
}{.}\hspace{.4pt}1111\discretionary{/}{%
}{/}j\hspace{.1pt}\discretionary{.}{%
}{.}\hspace{.4pt}1467\discretionary{%
}{-}{-}8659\hspace{.1pt}\discretionary{.}{%
}{.}\hspace{.4pt}2011\hspace{.1pt}\discretionary{.}{%
}{.}\hspace{.4pt}01923\hspace{.1pt}\discretionary{.}{%
}{.}\hspace{.4pt}x}}}


\bibitem{xu16semantic}
J.~Xu, Y.~Tao, and H.~Lin.
\newblock Semantic word cloud generation based on word embeddings.
\newblock In {\em Proc.\ of the IEEE Pacific Visualization Symposium
  (PacificVis)}, pp. 239--243. {IEEE}, 2016.
  \href{https://doi.org/10.1109/PacificVis.2016.7465278}
{doi: {{%
10\hspace{.1pt}\discretionary{.}{%
}{.}\hspace{.4pt}1109\discretionary{/}{%
}{/}PacificVis\hspace{.1pt}\discretionary{.}{%
}{.}\hspace{.4pt}2016\hspace{.1pt}\discretionary{.}{%
}{.}\hspace{.4pt}7465278}}}


\bibitem{yeo2008kim}
Z.~Yeo.
\newblock Emotional instant messaging with kim.
\newblock In {\em Extended Abstracts on Human Factors in Computing Systems
  (CHIEA)}, pp. 3729--3734. ACM, 2008.
  \href{https://doi.org/10.1145/1358628.1358921}
{doi: {{%
10\hspace{.1pt}\discretionary{.}{%
}{.}\hspace{.4pt}1145\discretionary{/}{%
}{/}1358628\hspace{.1pt}\discretionary{.}{%
}{.}\hspace{.4pt}1358921}}}


\bibitem{zhang2020DataQuilt}
J.~E. Zhang, N.~Sultanum, A.~Bezerianos, and F.~Chevalier.
\newblock {DataQuilt}: Extracting visual elements from images to craft
  pictorial visualizations.
\newblock In {\em Proc.\ of the ACM Conference on Human Factors in Computing
  Systems (CHI)}, pp. 45:1--45:13. ACM, 2020.
  \href{https://doi.org/10.1145/3313831.3376172}
{doi: {{%
10\hspace{.1pt}\discretionary{.}{%
}{.}\hspace{.4pt}1145\discretionary{/}{%
}{/}3313831\hspace{.1pt}\discretionary{.}{%
}{.}\hspace{.4pt}3376172}}}


\bibitem{Zou16calligram}
C.~Zou, J.~Cao, W.~Ranaweera, I.~Alhashim, P.~Tan, A.~Sheffer, and H.~Zhang.
\newblock Legible compact calligrams.
\newblock {\em ACM Trans. Graph.}, 35(4):122:1--122:12, 2016.
  \href{https://doi.org/10.1145/2897824.2925887}
{doi: {{%
10\hspace{.1pt}\discretionary{.}{%
}{.}\hspace{.4pt}1145\discretionary{/}{%
}{/}2897824\hspace{.1pt}\discretionary{.}{%
}{.}\hspace{.4pt}2925887}}}


\end{thebibliography}
%



%




\end{document}